\documentclass[twocolumn,amsfonts,showpacs,superscriptaddress,nofootinbib]{revtex4-1}
\usepackage[pdftex,breaklinks=true,colorlinks=true]{hyperref} 
\usepackage[dvipdfmx]{graphicx}
\usepackage{amssymb,amsmath,amsthm,booktabs,mathtools}
\usepackage{bbm}
\usepackage{bm}
\usepackage{color}
\usepackage{tikz}
\usepackage{pgfplots}
\usepackage{enumitem}
\usepackage{hyperref}
\newcommand{\la}{\langle}\newcommand{\ra}{\rangle}
\newcommand{\ket}[1]{\left|#1\right>}
\newcommand{\bra}[1]{\left<#1\right|}
\newcommand{\ii}{\mathrm{i}}
\newcommand{\n}{\nonumber \\}
\newcommand{\dd}{{\rm d}}

\newcommand{\tgamma}{\Tilde{\Gamma}}

\usepackage{xcolor}

\begin{document}
\title{Non-Hermitian quantum impurity systems in and out of equilibrium: noninteracting case}
\author{Takato Yoshimura}
\affiliation{Institut de Physique Th\'eorique Philippe Meyer, \'Ecole Normale Sup\'erieure, \\ PSL University, Sorbonne Universit\'es, CNRS, 75005 Paris, France}
\affiliation{Department of Mathematics, King's College London, Strand, London WC2R 2LS, U.K.}

\author{Kemal Bidzhiev}
\affiliation{
Université Paris-Saclay, CNRS, LPTMS, 91405, Orsay, France}
\author{Hubert Saleur}
\affiliation{Institut de Physique Th\'eorique, Universit\'e Paris Saclay, CEA, CNRS, F-91191 Gif-sur-Yvette, France}

\date{\today}

\begin{abstract}
 We provide systematic analysis on a non-Hermitian $\mathcal{PT}$-symmetric quantum impurity system both in and out of equilibrium, based on exact computations. In order to understand the interplay between non-Hermiticity and Kondo physics, we focus on a prototypical noninteracting impurity system, the resonant level model, with complex coupling constants. Explicitly constructing biorthogonal basis, we study its thermodynamic properties as well as the Loschmidt echo starting from the initially disconnected two free fermion chains. Remarkably, we observe the universal crossover physics in the Loschmidt echo, both in the $\mathcal{PT}$ broken and unbroken regimes. We also find that the ground state quantities we compute in the $\mathcal{PT}$ broken regime can be obtained by analytic continuation. It turns out that Kondo screening ceases to exist in the $\mathcal{PT}$ broken regime, which was also previously predicted in the non-hermitian Kondo model. All the analytical results are corroborated against biorthogonal free fermion numerics.
\end{abstract}
\maketitle

\section{Introduction}
 Quantum impurity systems, such as the Kondo model, serve as representative examples where nonperturbative quantum many-body effects give rise to unconventional low-temperature behaviours that are in stark contrast with those of Fermi liquids \cite{hewson_1993}. One of the cornerstones of such impurity models is the screening of the impurity spin at sufficiently low temperature (lower than the typical temperature scale called Kondo temperature $T_K$), which is characterised by the spin singlet made of the impurity spin and low-lying excitations \cite{PhysRevB.53.9153}. The impurity spin is therefore deactivated by healing with the lead baths, thereby the typical low temperature behaviour of the spin susceptibility $\chi(T)\sim1/T$ is replaced by the approach towards a constant value $\chi(T)\to\chi_0$ as $T\to0$. A great deal of efforts has been made to understand the nature of Kondo screening in the past decades making use of renormalisation group \cite{Anderson_1970}, Bethe ansatz\cite{PhysRevLett.45.379,doi:10.1080/00018738300101581}, and numerical methods such as the numerical renormalisation group (NRG) \cite{RevModPhys.80.395} and density-matrix renormalisation group (DMRG) \cite{PhysRevLett.101.140601}. Recently, experimental investigations of the Kondo physics have been put forward using alkaline-earth atoms, opening up an unprecedented avenue to detect the Kondo cloud, which has remained experimentally elusive \cite{PhysRevB.97.155156,PhysRevLett.118.196803}.
 
 There recently has been an interest in exploring what happens to Kondo physics in the  non-hermitian case \cite{ptkondo,lourenptkondo6}.   This is partly motivated by set-ups such as the one in \cite{ptkondo},  where a non-Hermitian Kondo model is obtained as an effective model that characterizes ultra-cold mobile and immobile atoms that undergo inelastic scatterings (resulting in two-body losses). Another motivation comes from the old observation \cite{doi:10.1142/S0217751X93002265,ZAMOLODCHIKOV1994436,SALEUR1994205} that making the couplings complex in the Kosterlitz-Thouless (KT) renormalization group (RG) flow can profoundly change the low-energy physics, and lead sometimes to ``circular'' behavior. This RG flow is relevant for the sine-Gordon model with complex coupling constant, a  model that found early applications in 2d statistical mechanics (e.g. polymers), and was also studied more recently in  \cite{Ashida2017}  in the context of dissipative quantum mechanics.  The KT RG flow is of course also relevant to  the(anisotropic) Kondo model, and the circular behavior was explored in this context in \cite{lourenptkondo6}, where it was concluded  that the Kondo effect might disappear in the non-hermitian case. 
 
 Non-Hermitian ``hamiltonians'' of physical relevance are usually $\mathcal{PT}$-symmetric, and can  exhibit both $\mathcal{PT}$-broken and unbroken  regimes. The spectrum of a  system with unbroken $\mathcal{PT}$-symmetry is actually real, and such systems  can essentially be treated like  Hermitian ones. The situation is  different in the $\mathcal{PT}$-broken regime where generically eigenstates with complex eigenvalues are present. The new low-energy behaviour in non-hermitian versions of the sine-Gordon model or the Kondo model is expected to occur in such broken symmetry regimes. 
 
 Since the physics of interest in this context involves typically many-body effects, available methods to understand what is going on are few, apart from perturbative RG calculations. Attempts using the Bethe-ansatz \cite{doi:10.1142/S0217751X93002265,ZAMOLODCHIKOV1994436,SALEUR1994205} have in particular been much less successful than in the usual, Hermitian case, due in part to the appearance of bound-states of complex or even purely imaginary energy (``monstrons'' \cite{ZAMOLODCHIKOV1994436}). Reference \cite{ptkondo} does  partly explore the issue in the Kondo case, concluding that the Kondo effect might disappear in the broken $\mathcal{PT}$-symmetry regime; the calculations in this reference are  however only a starting point, and leave many questions open.

 In the Hermitian case, a great deal of the physics of quantum impurity problems both in and out of equilibrium can be learned without the Bethe-ansatz, by studying the  resonant level model (RLM) \cite{Ghosh_2014}, which occurs as a simpler (Toulouse) limit of interacting impurity systems such as the anisotropic Kondo model and the interacting resonant level model. 
 
It is thus natural, in order to clarify what happens  to quantum-impurity problems in the  non-Hermitian case, to start by studying a  $\mathcal{PT}$-symmetric version of the RLM (PTRLM) model. The fact that the model is essentially free will allow us to carefully analyse the issues of biorthogonal bases, bound states and $\mathcal{PT}$-symmetry breaking, paving the way for further studies of the more delicate, interacting case. We will see that the conclusions of earlier studies must be taken with a grain of salt, and that there is probably a lot of physics left to be understood in this problem.

The paper is organised as follows. The spectrum of the model is obtained in section II, where we identify the different regimes of the model, and the $\mathcal{PT}$-symmetry breaking.  The dot occupancy is studied in section III, and the boundary free-energy in section IV.  These two equilibrium quantities are known to exhibit characteristic features of Kondo physics in the hermitian case. While we observe similar features in the $\mathcal{PT}$-symmetric regime, very different properties occur when that symmetry is broken. Section V deals with non-equilibrium physics, where we look for another angle to understand the physics of the broken 
$\mathcal{PT}$-symmetric phase by studying the Loschmidt echo. Consideration of analytic continuation are discussed in section VI, where we suggest another, more direct way to handle the problem that could be generalized to the interacting case. Further observations and conclusions are discussed in the conclusion.

We finally mention that our model can be in principle realised using alkaline-earth atoms with a tunable $\mathcal{PT}$ symmetric optical lattice \cite{PhysRevB.97.155156,PhysRevLett.111.215304,PhysRevLett.120.143601}.

\section{Spectrum of the $\mathcal{PT}$-symmetric Resonant level model}
\subsection{The model}
The Hamiltonian of the $\mathcal{PT}$-symmetric RLM model (PTRLM) we study takes the following form:
\begin{align}\label{eq:ptrlm}
H&=H_{\rm A}+H_{\rm B}+H_{\rm d}\n
H_{\rm A}&=-t\sum_{x=-N/2}^{-2}(c^\dagger_xc_{x+1}+{\rm h.c.})\n 
H_{\rm B}&=-t\sum_{x=1}^{N/2-1}(c^\dagger_xc_{x+1}+{\rm h.c.}) \n
H_\mathrm{d}&= -\gamma(c^\dagger_{-1}c_0+{\rm h.c.})-\gamma^*(c^\dagger_{1}c_0+{\rm h.c.}) \n
&= -J e^{i\varphi}(c^\dagger_{-1}c_0+{\rm h.c.})-J e^{-i\varphi}(c^\dagger_{1}c_0+{\rm h.c.}) ,
\end{align}
where  $H_{\rm A/B}$ describes the kinetic energy of free spinless fermions in the left and right wires, $H_{\rm d}$ defines tunneling from the wires to the dot impurity with a hopping amplitude $\gamma=\gamma_1+\ii\gamma_2 =Je^{i \varphi}$. In what follows, we set tunneling strength between wires $t=1$ for brevity. That this model is $\mathcal{PT}$-symmetric follows from the following definitions \cite{kawabatatopcond,lourenptkondo6} of how $\mathcal{P}$ and $\mathcal{T}$ act on operators
\begin{eqnarray}
   \hat{\mathcal{P}}c_x\hat{\mathcal{P}}^{-1}=c_{-x},\quad \hat{\mathcal{T}}c_x\hat{\mathcal{T}}^{-1}&=&c_x,\quad \hat{\mathcal{T}}\ii\hat{\mathcal{T}}^{-1}=-\ii. \\
   (\mathcal{\hat{P}\hat{T}})  H   (\mathcal{\hat{P}\hat{T}}) &=& H
\end{eqnarray}
\begin{figure}[h]
\includegraphics[width=0.45\textwidth]{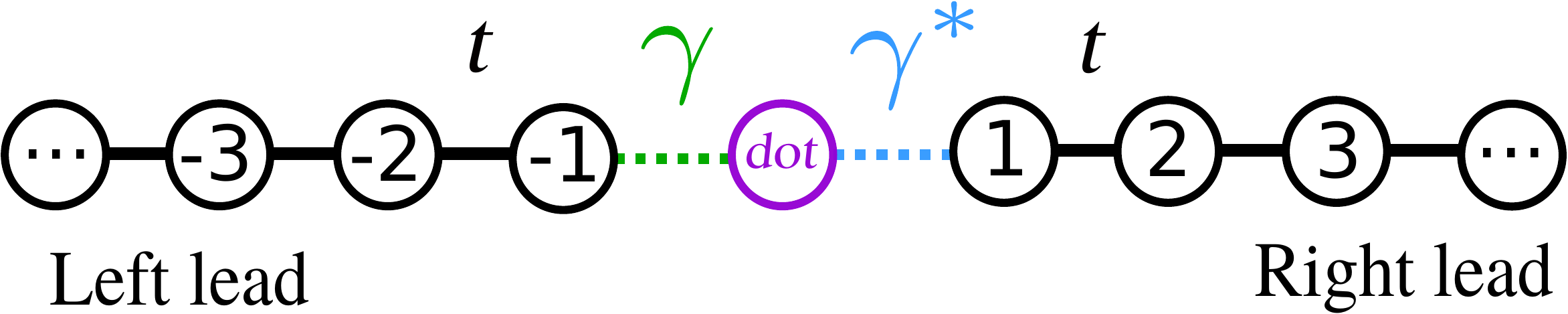}
\caption{A schematics of the $\mathcal{PT}$ resonant level model \eqref{eq:ptrlm}. The impurity couples to the leads with hopping amplitudes $\gamma$ and $\gamma^*$ (or tunneling strength).}
\label{ptrlmschem}
\end{figure}

\begin{figure}[h]
\includegraphics[width=.5\textwidth]{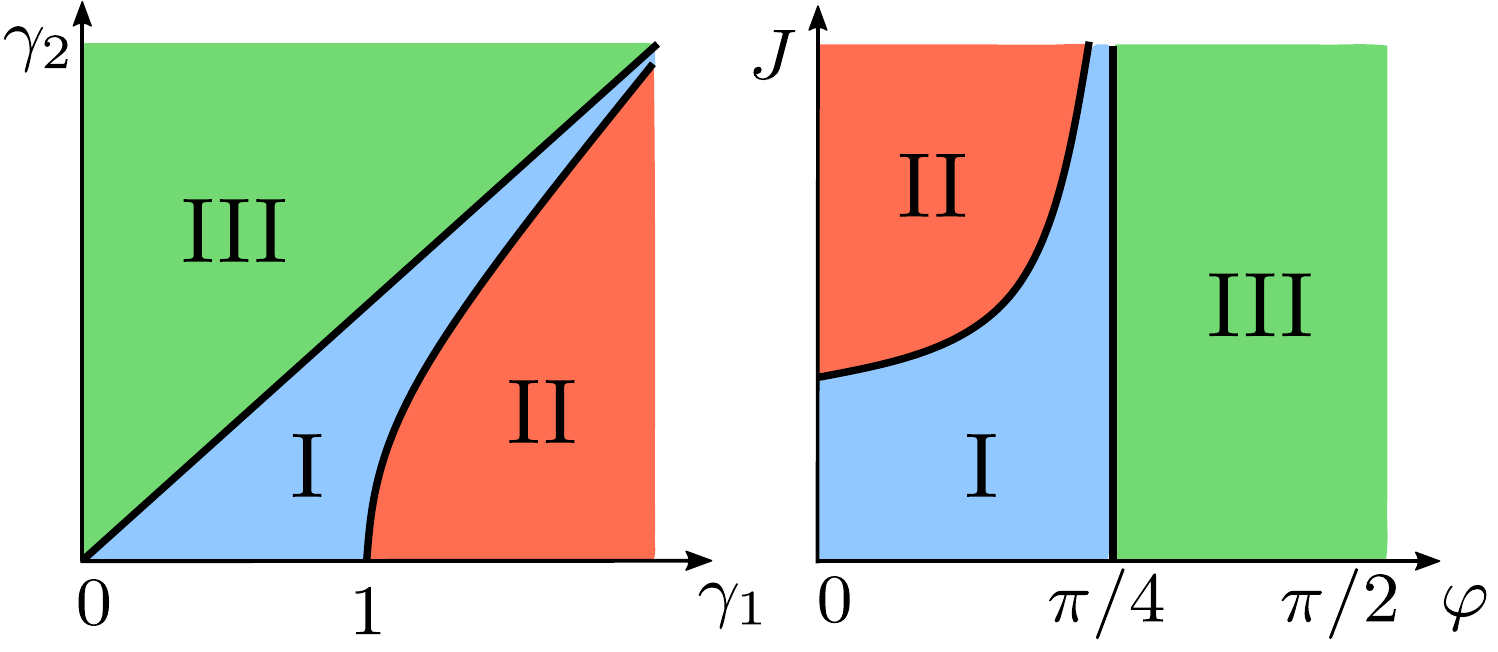}
\caption{The phase diagram of the PTRLM for different coordinates $\gamma = \gamma_1+ i \gamma_2 =J e^{i \varphi}$. The phase I and II are the $\mathcal{PT}$-unbroken phases that are distinguished by the presence (phase II) and the absence (phase I) of bound states. The phase III corresponds to $\mathcal{PT}$-broken phase with complex-valued single particle energies.}
\label{fig:phasediagram}
\end{figure}

\subsection{The eigenstates}

In order to study the spectrum of $\mathcal{PT}$-RLM \eqref{eq:ptrlm}, let us start with solving the eigenvalue problems of the model by constructing the biorthogonal basis \cite{biorth}. We write a generic one-particle {\it right} eigenstate with the eigenvalue $E$ as
\begin{equation}
|R\rangle=\sum_xR_{E}(x)|x\rangle.
\end{equation}
where $|x\rangle=c^\dagger_x|0\rangle$. First, for $1<|x|<N/2$, $R_{E}(x)$ solves
\begin{equation}\label{eigen1}
R_{E}(x-1)+R_{E}(x+1)=-ER_{E}(x).
\end{equation}
Let us take the following ansatz~\cite{Landau1981Quantum} for $x\neq0$,
\begin{equation}\label{righteigfun}
R(k;x)=\begin{dcases*}
    \mathcal{N}_+(k)(e^{\ii kx}+\psi_+(k)e^{-\ii kx}) & $x>0$\\
        \mathcal{N}_-(k)(e^{\ii kx}+\psi_-(k)e^{-\ii kx}) & $x<0$
\end{dcases*},
\end{equation}
where $\psi_+=e^{2\ii k\delta(k)}$, $\psi_-=e^{-2\ii k\delta(k)}$, and $\mathcal{N}_\pm(k)$ are normalisations. This solves \eqref{eigen1} where the quasi-momentum $k$ is related to $E_k$ by the dispersion relation $E_k=-2\cos k$. The phase shift $\delta(k)$ is determined by
\begin{align}
R(k;2)+\gamma^*R(k;0)&=-E_kR(k;1) \label{erel1}\\
\gamma R(k;-1)+\gamma^*R(k;1)&=-E_kR(k;0)\label{erel2}\\
R(k;-2)+\gamma R(k;0)&=-E_kR(k;-1)\label{erel3}
\end{align}
Solving these self-consistently, we find that, in fact, there are two types of solutions that are labeled by $\sigma=\pm1$. The phase shifts corresponding to these solutions accordingly depend on $\sigma$, and satisfy
\begin{equation}\label{ps}
e^{2\ii k\delta_\sigma(k)}=\frac{1-\Gamma_\sigma e^{2\ii k}}{\Gamma_\sigma-e^{2\ii k}},
\end{equation}
where $\Gamma_\sigma=(1+\sigma)\Tilde{\Gamma}-1$ with $\Tilde{\Gamma}=(\gamma^2+(\gamma^*)^2)/2=J^2\cos2\varphi$. Note that $\tgamma=J^2$ in the Hermitian case. Defining $\mathcal{N}_{\mathrm{R},+}=2\mathcal{N}_+e^{\ii k\delta_+(k)}$ and $\mathcal{N}_{\mathrm{R},-}=2\ii\mathcal{N}_+e^{\ii k\delta_-(k)}$, it is then a simple matter to observe that the wave function $R_\sigma(k;x)$ takes the following form
\begin{equation}\label{righteigenfun1}
    R_\sigma(k;x)=\begin{dcases*}
    \mathcal{N}_{{\rm R},\sigma}(k)\cos[k(x-\delta_\sigma(k))] & $x>0$\\
        \mathcal{M}_\sigma\mathcal{N}_{{\rm R,\sigma}}(k)\cos[k(x+\delta_\sigma(k))] & $x<0$
\end{dcases*},
\end{equation}
where $\mathcal{M}_+=\gamma^2/|\gamma|^2$ and $\mathcal{M}_-=-(\gamma^*)^2/|\gamma|^2$. Note that $R_\sigma(k;0)$ is is given by \eqref{erel2}.
Finally, the quantisation condition is obtained from the equation for $x=N/2$
\begin{equation}\label{quantcond}
\cos k(N/2+1-\delta_\sigma(k))=0,
\end{equation}
which, together with \eqref{ps}, gives
\begin{equation}\label{ps2}
e^{\ii k N}=\frac{\Gamma_\sigma-e^{-2\ii k}}{\Gamma_\sigma-e^{2\ii k}}.
\end{equation}
The construction of the left eigenstate $|L\rangle=\sum_xL_{E}(x)|x\rangle$ can also be carried out in the analogous way. For real $k$, it reads
\begin{equation}\label{lefteigenfun1}
    L_\sigma(k;x)=\begin{dcases*}
    \mathcal{N}_{{\rm L,\sigma}}(k)\cos[k(x-\delta_\sigma(k))] & $x>0$\\
        \mathcal{M}^*_\sigma\mathcal{N}_{{\rm L,\sigma}}(k)\cos[k(x+\delta_\sigma(k))] & $x<0$
\end{dcases*}.
\end{equation}
\subsection{Bound states}
So far we have focused on the case where $k$ is real, but we can also find solutions of the equation \eqref{ps2} with complex $k$, corresponding to bound (localised) states. In general it is hard to determine the precise form of their eigenfunctions except when thermodynamic limit ($N\to\infty$) is taken, which is the case we shall consider below. It turns out that there are three distinct regimes which we present in Fig. \ref{fig:phasediagram}. In what follows, without loss of generality, we shall focus on $\gamma_1,\gamma_2>0$.

$\bullet$ Phase I ($\gamma_2<\gamma_1<\sqrt{1+\gamma^2_2}$): $\mathcal{PT}$-unbroken regime without bound states.\\
No complex $k$ is allowed in this phase, hence there is no bound state and the excitation is solely characterized by the real $k$. The right and left eigenfunctions are given by \eqref{righteigenfun1} and \eqref{lefteigenfun1}. The normalization condition for the left and right wavefunctions reads
\begin{align}
    \mathcal{N}_{{\rm L},+}(k)^*\mathcal{N}_{{\rm R,+}}(k)&=\frac{(\gamma^*)^2}{\gamma^2+(\gamma^*)^2}\Big[\frac{N}{4}+\frac{1-\Gamma_+\cos2k}{\Delta_+(k)}\Big]^{-1} \n
    \mathcal{N}_{{\rm L},-}(k)^*\mathcal{N}_{{\rm R,-}}(k)&=\frac{\gamma^2}{\gamma^2+(\gamma^*)^2}\Big[\frac{N}{4}+\frac{1}{2}\Big]^{-1}
\end{align}
where $\Delta_+(k)=1-2\Gamma_+\cos2k+\Gamma_+^2=4(\Tilde{\Gamma}^2+(1-2\Tilde{\Gamma})\cos^2k)$.

\begin{figure}[h]
\includegraphics[width=0.5\textwidth]{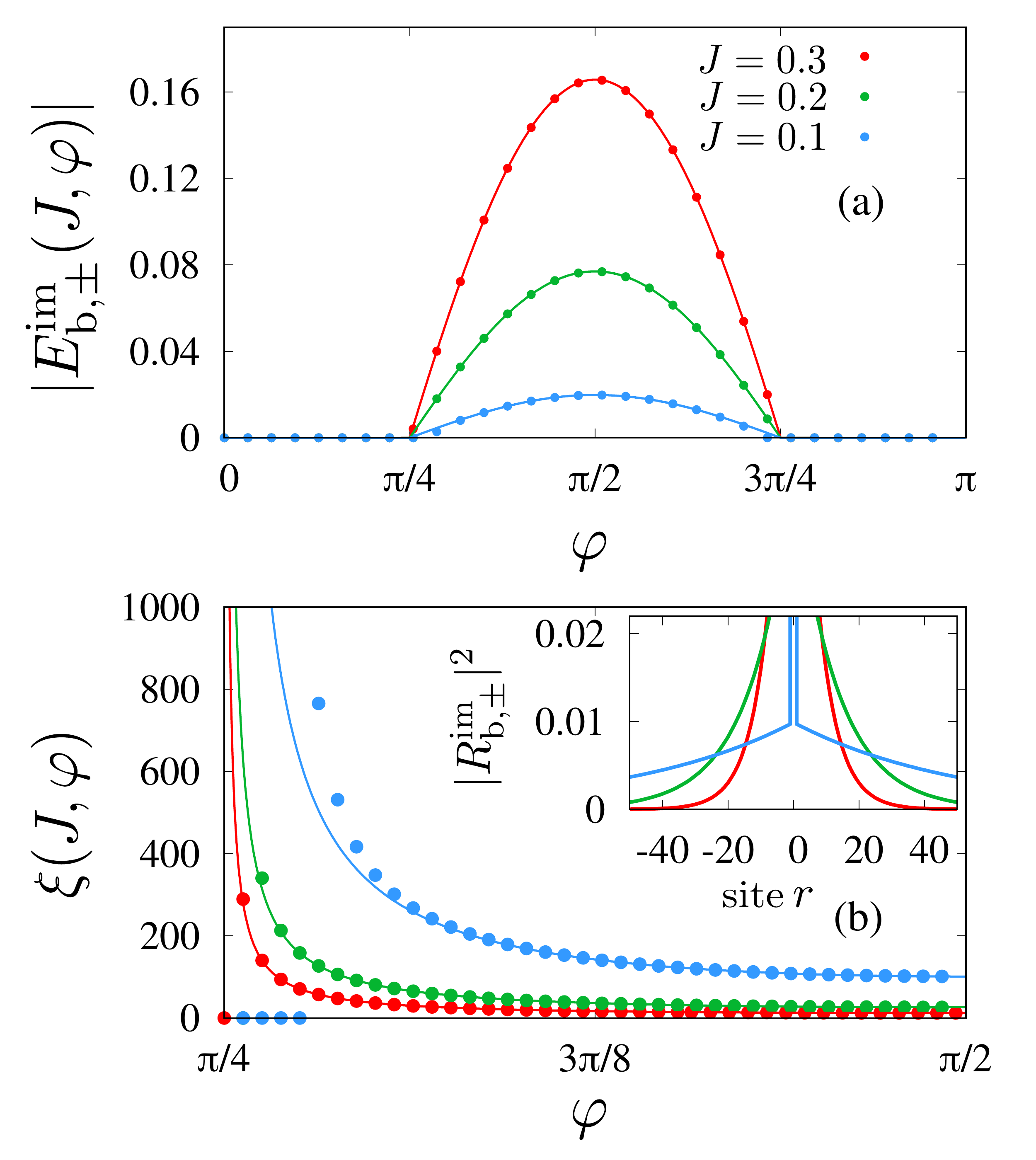}
\caption{Imaginary bound state energy $E^{\rm im}_{\rm b,\pm}$ Eq.\eqref{eq:imEb} (panel a)  and a localization length $\xi$ (panel b) as a function of a coupling argument $\varphi ( \gamma=J e^{i \varphi})$ for different values of a coupling magnitude $J$. Inset in panel (b): $|\psi|^2$ of an imaginary bound state for $\varphi = \pi/2$.
Lines represent Eq.\eqref{eq:imEb}(a) and Eq.(\ref{leigimagbound}, \ref{reigimagbound})(b), while points label numerical results.
Localization length $\xi$ diverges and becomes larger than system size $N$ in the vicinity of $\varphi=\pi/4$ and, therefore, imaginary bound states with $\xi > N$ are absent (here $N=1025$).
}
\label{fig:En_Loc}
\end{figure}

$\bullet$ Phase II ($\gamma_1>\sqrt{1+\gamma^2_2}$): $\mathcal{PT}$-unbroken regime with bound states.\\
In this phase, on top of the real eigenvalues as in the phase I, bound states being localised around the dot can also be formed. The corresponding right and left eigenfunctions are
\begin{equation}\label{reigreagbound}
    R^\mathrm{re}_{\mathrm{b},\pm}(x)=\begin{dcases*}
    \mathcal{N}_{\mathrm{R},\mathrm{b}}^{\rm re}(\pm1)^xe^{-x/\xi} & $x>0$\\
    \mathcal{N}_{\mathrm{R},\mathrm{b}}^{\rm re}\frac{1}{\gamma^*} & $x=0$\\
        \frac{\gamma^2}{|\gamma|^2}\mathcal{N}_{\mathrm{R},\mathrm{b}}^{\rm re}(\pm1)^xe^{x/\xi} & $x<0$
\end{dcases*},
\end{equation}
\begin{equation}\label{leigreagbound}
    L^\mathrm{re}_{\mathrm{b},\pm}(x)=\begin{dcases*}
    \mathcal{N}_{\mathrm{L},\mathrm{b}}^{\rm re}(\mp1)^xe^{-x/\xi} & $x>0$\\
    \mathcal{N}_{\mathrm{L},\mathrm{b}}^{\rm re}\frac{1}{\gamma} & $x=0$\\
        \frac{(\gamma^*)^2}{|\gamma|^2}\mathcal{N}_{\mathrm{L},\mathrm{b}}^{\rm re}(\mp1)^xe^{x/\xi} & $x<0$
\end{dcases*},
\end{equation}
where $\xi=2/\log|\Gamma_+|=2/\log|\gamma^2+(\gamma^*)^2-1|$ is the localization length of the bound states. Associated eigenvalues and the normalization condition for $\mathcal{N}_{\mathrm{R},\mathrm{b}}^{\rm re}$ and $\mathcal{N}_{\mathrm{L},\mathrm{b}}^{\rm re}$ are provided by
\begin{equation}
    E^\mathrm{re}_{\mathrm{b},\pm}=\pm2\frac{\gamma^2_1-\gamma^2_2}{\sqrt{2\gamma^2_1-2\gamma^2_2-1}}=\pm\frac{\gamma^2+(\gamma^*)^2}{\sqrt{\gamma^2+(\gamma^*)^2-1}}
\end{equation}
and
\begin{equation}
    (\mathcal{N}_{\mathrm{L},\mathrm{b}}^{\rm re})^*\mathcal{N}_{\mathrm{R},\mathrm{b}}^{\rm re}=\frac{(\gamma^*)^2(\gamma^2+(\gamma^*)^2-2)}{2(\gamma^2+(\gamma^*)^2-1)}.
\end{equation}
$\bullet$ Phase III ($\gamma_2>\gamma_1$): $\mathcal{PT}$-broken regime with bound states.\\
In this regime, $\mathcal{PT}$-symmetry is spontaneously broken, and the bound states with imaginary eigenvalues emerge. Eigenmodes that give these states are $k_b=\frac{\ii}{2}\log(-1)+\frac{\ii}{2}\log|\Gamma|+n\pi$ for $n\in\mathbb{Z}$. Due to the two possible branches of $\log(-1)$, we have two right eigenfunctions and associated left eigenfunctions
\begin{equation}\label{reigimagbound}
    R^\mathrm{im}_{\mathrm{b},\pm}(x)=\begin{dcases*}
    \mathcal{N}_{\mathrm{R},\mathrm{b}}^{\rm im}(\pm\ii)^xe^{-x/\xi} & $x>0$\\
    \mathcal{N}_{\mathrm{R},\mathrm{b}}^{\rm im}\frac{1}{\gamma^*} & $x=0$\\
        \frac{\gamma^2}{|\gamma|^2}\mathcal{N}_{\mathrm{R},\mathrm{b}}^{\rm im}(\mp\ii)^xe^{x/\xi} & $x<0$
\end{dcases*},
\end{equation}
\begin{equation}\label{leigimagbound}
    L^\mathrm{im}_{\mathrm{b},\pm}(x)=\begin{dcases*}
    \mathcal{N}_{\mathrm{L},\mathrm{b}}^{\rm im}(\mp\ii)^xe^{-x/\xi} & $x>0$\\
     \mathcal{N}_{\mathrm{L},\mathrm{b}}^{\rm im}\frac{1}{\gamma} & $x=0$\\
        \frac{(\gamma^*)^2}{|\gamma|^2}\mathcal{N}_{\mathrm{L},\mathrm{b}}^{\rm im}(\pm\ii)^xe^{x/\xi} & $x<0$
\end{dcases*},
\end{equation}
 Eigenvalues of these imaginary bound states are (see Fig.\ref{fig:En_Loc})
\begin{equation}
    E^\mathrm{im}_{\mathrm{b},\pm}=\mp\ii\frac{\gamma^2+(\gamma^*)^2}{\sqrt{1-\gamma^2-(\gamma^*)^2}}=\mp2\ii\frac{\tgamma}{\sqrt{1-2\tgamma}}.
\label{eq:imEb}
\end{equation}

Later, we shall also denote the absolute value of the bound state eigenenergies, whether real or imaginary, as $E_\mathrm{b}=2|\tgamma|/\sqrt{|1-2\tgamma}|$. It is readily seen that the right and left eigenfunctions are indeed orthogonal with the normalization condition
\begin{equation}
    (\mathcal{N}_{\mathrm{L},\mathrm{b}}^{\rm im})^*\mathcal{N}_{\mathrm{R},\mathrm{b}}^{\rm im}=\frac{(\gamma^*)^2(2-\gamma^2-(\gamma^*)^2)}{2(1-\gamma^2-(\gamma^*)^2)}.
\end{equation}
This observation implies that, in our $\mathcal{PT}$-symmetric RLM \eqref{eq:ptrlm}, the spontaneous breaking of $\mathcal{PT}$-symmetriy is solely caused by the presence of bound states that are localized across the impurity. Furthermore, as $\gamma_1$ and $\gamma_2$ approach, the two states characterised by $R^\mathrm{im}_{\mathrm{b},\pm}(x)$ coalesce and form exponential points with diverging localisation length $\xi$. In general eigenenergies with positive imaginary values inevitably give rise to exponential growth of physical quantities, such as correlation functions, in time evolution. This is obviously incompatible with unitarity and causality that are required for any physically sensible systems. The existence of such anomalous excitations in non-Hermitian systems was actually proposed decades ago in the imaginary Sine-Gordon model \cite{doi:10.1142/S0217751X93002265,ZAMOLODCHIKOV1994436,SALEUR1994205}, and coined {\it monstrons}. Such exponential divergency could be, however, canceled out by normalising the time-evolved state $|\psi(t)\rangle$ by its biorthogonal norm $\langle\Tilde{\psi}(t)|\psi(t)\rangle$, where $\langle\Tilde{\psi}(t)|$ is the {\it associated} state of $|\psi(t)\rangle$ (see Appendix \ref{biortho} for the discussion about the associated state and the biorthogonal norm).

We finally note that, our approach does not work on the line of exceptional points $\gamma_1=\gamma_2$ where the Hamiltonian \eqref{eq:ptrlm} is {\it not} diagonalizable, and accordingly the wave function is not normalizable \cite{Mirieaar7709}. One therefore needs to devise an alternative method to study the spectrum at these particular points.
\section{Dot occupancy and magnetization}
\subsection{Diagonalized Hamiltonian and the ground state}
Let us first define mode operators $c^\dagger_{R,k}$ an$c^\dagger_{L,k}$ for real $k$ as
\begin{equation}
c^\dagger_{R,k\sigma}=\sum_{x}R_\sigma(k;x)c^\dagger_x,\quad c^\dagger_{L,k\sigma}=\sum_{x}L_\sigma(k;x)c^\dagger_x,
\end{equation}
and for complex $k$, as
\begin{equation}
c^\dagger_{R,\epsilon}=\sum_{x}R_{\mathrm{b},\epsilon}(x)c^\dagger_x,\quad c^\dagger_{L,\epsilon}=\sum_{x}L_{\mathrm{b},\epsilon}(x)c^\dagger_x,
\end{equation}
so that $|R_\sigma(k)\rangle=c^\dagger_{R,k}|0\rangle$ and $|L_{\sigma}(k)\rangle=c^\dagger_{L,k}|0\rangle$, and $|R_{\mathrm{b},\epsilon}\rangle=c^\dagger_{R,\epsilon}|0\rangle$ and $|L_{\mathrm{b},\epsilon}\rangle=c^\dagger_{L,\epsilon}|0\rangle$. Using the orthogonality relation $\int{\rm d} xL^*_\sigma(k;x)R_{\sigma'}(k';x)=\delta_{k,k'}\delta_{\sigma,\sigma'}$, these are equivalently expressed as
\begin{equation}
c^\dagger_x=\sum_{k,\sigma}L_\sigma^*(k;x)c^\dagger_{R,k\sigma},\quad c_x=\sum_{k,\sigma}L_\sigma(k;x)c_{R,k\sigma}
\end{equation}
or
\begin{equation}
c^\dagger_x=\sum_{k,\sigma}R_\sigma^*(k;x)c^\dagger_{L,k\sigma},\quad c_x=\sum_{k,\sigma}R_\sigma(k;x)c_{L,k\sigma},
\end{equation}
where summation is taken over both real and complex $k$ \footnote{It is important to note that actually $k$ depends on $\sigma$ through \eqref{ps2}, hence the summation over $k$ and $\sigma$ are not independent. Here, for ease of notation we decided not to explicitly show this condition in writing the summations.}.
Notice that $c^\dagger_{R,k\sigma}$ and $c_{L,k\sigma}$ satisfy the anticommutation relation $\{c^\dagger_{R,k\sigma},c_{L,k'\sigma'}\}=\delta_{k,k'}\delta_{\sigma,\sigma'}$. In terms of these mode operators, \eqref{eq:ptrlm} can be diagonalized as
\begin{equation}
H=\sum_{k,\sigma}E_kc^\dagger_{R,k\sigma}c_{L,k\sigma}+\sum_{\epsilon=\pm}E_{\mathrm{b},\epsilon}c^\dagger_{R,\epsilon}c_{L,\epsilon},
\end{equation}
where the second term accounts for the possible bound states that appear in the phase II and III (in the phase I this term vanishes) with $E_{\mathrm{b},\epsilon}$ being either $E^\mathrm{re}_{\mathrm{b},\epsilon}$ or $E^\mathrm{im}_{\mathrm{b},\epsilon}$. We shall construct the ground state of PTRLM as in the hermitian case, i.e. by filling modes that lower the real part of the energies. When $\mathcal{PT}$ symmetry is not broken, there is no ambiguity in defining the ground state as per the above principle, hence the left and the right ground state state can be built as
\begin{align}
    |\mathrm{GS}\rangle_R&=\prod_\sigma\prod_{k\in\mathrm{Fermi\ sea}}c^\dagger_{R,k\sigma}|0\rangle, \\
    |\mathrm{GS}\rangle_L&=\prod_\sigma\prod_{k\in\mathrm{Fermi\ sea}}c^\dagger_{L,k\sigma}|0\rangle.
\end{align}
Notice that the ground state can be uniquely constructed even if $\mathcal{PT}$ symmetry is spontaneously broken (i.e. phase III), except when the system is at half-filling. At half-filling ($\mu=0$), the real part of the bound states lies on the boundary of the Fermi sea, hence it could be either filled or not filled. In this article, we {\it define} the ground state in the phase III at half-filling by filling those two bound states so as to avoid complex ground state energies. We stress that there is no fundamental reason why we have to do so; we might as well fill just one of them at the price of having the ground state with complex energy. Our claim here is that our way of constructing the ground state is physically sensible (no complex ground state energy), and also gives rise to the behaviour of the dot density that is consistent with a previous literature \cite{ptkondo}. Furthermore, the analytic continuation to the dot density, which we will elaborate in section \ref{anacont}, works only with this choice of the ground state.  
\subsection{Dot density and susceptibility}\label{dotsuscep}
One of the simplest yet nontrivial objects in quantum impurity systems is the dot density. Let us first compute it when $\mathcal{PT}$-symmetry is not spontaneously broken. It is defined by

\begin{align}\label{dot1}
d&=\mathrm{Tr}(\varrho c^\dagger_0c_0) \n
&=\sum_k\frac{L_+^*(k;0)R_+(k;0)}{1+e^{\beta(E_k-\mu)}}+\sum_\epsilon\frac{L^*_{\mathrm{b},\epsilon}(0)R_{\mathrm{b},\epsilon}(0)}{1+e^{\beta(E_{\mathrm{b},\epsilon}-\mu)}}.
\end{align}
The second term in \eqref{dot1} vanishes in the phase I, and $d$ reads simply, in the thermodynamic limit,
\begin{equation}
    \int_{-\pi}^\pi\frac{\dd k}{2\pi}\frac{1}{1+e^{\beta(E_k-\mu)}}\frac{\Tilde{\Gamma}\sin^2k}{\Tilde{\Gamma}^2+(1-2\Tilde{\Gamma})\cos^2k}. \label{dot2}
\end{equation}
If we take the real value limit $\gamma_2=0$, this reproduces the known result in the real RLM. In the phase II, we have contributions from the bound states, which can be calculated as
\begin{align}
    d^\mathrm{re}_\mathrm{bound}&=\sum_\epsilon\frac{L^*_{\mathrm{b},\epsilon}(0)R_{b,\epsilon}(0)}{1+e^{\beta(E^\mathrm{re}_{\mathrm{b},\epsilon}-\mu)}}\n
    &=\frac{2-\gamma^2-(\gamma^*)^2}{1-\gamma^2-(\gamma^*)^2}\n 
    &\quad\times\frac{1+e^{-\beta\mu}\cosh \beta E_\mathrm{b}}{1+2e^{-\beta\mu}\cosh\beta E_\mathrm{b}+e^{-2\beta\mu}}, 
\end{align}
where we recall $E_\mathrm{b}=|E^\mathrm{re}_{\mathrm{b},\pm}|$.
The dot density in the phase II is then given by
\begin{equation}\label{dot3}
    d=\int_{-\pi}^\pi\frac{\dd k}{2\pi}\frac{1}{1+e^{\beta(E_k-\mu)}}\frac{\Tilde{\Gamma}\sin^2k}{\Tilde{\Gamma}^2+(1-2\Tilde{\Gamma})\cos^2k}+d^{\mathrm{re}}_\mathrm{bound}.
\end{equation}
Note that the dot densities in both phase I and II are identical to 1/2 when $\mu=0$ due to the particle-hole symmetry (see Appendix \ref{dotdensity} for the proof).

Next, we turn our attention to the phase III where $\mathcal{PT}$-symmetry is broken by the presence of complex eigenstates. It turns out that thermodynamics is ill-defined in this phase: any thermal state characterised by a Gibbs ensemble is {\it dynamically unstable}. Concretely speaking, a Gibbs measure $\varrho=\exp(-\beta(H-\mu Q))/Z$ is not invariant under time evolution due to the breakdown of  Heisenberg picture (See the discussion in the appendix \ref{biortho}). Typically it is expected that stationary states in this phase are not universal and strongly depending on the initial state. Nonetheless, the average of an arbitrary observable in the ground state is still stationary and therefore makes physical sense. Then, we can calculate the dot density as
\begin{equation}
    d=\int_{-k_\mathrm{F}}^{k_\mathrm{F}}\frac{\dd k}{2\pi}\frac{\Tilde{\Gamma}\sin^2k}{\Tilde{\Gamma}^2+(1-2\Tilde{\Gamma})\cos^2k}+\frac{2-\gamma^2-(\gamma^*)^2}{1-\gamma^2-(\gamma^*)^2}.
\end{equation}
Note that the appearance of the second term can amount to the dot density larger than 1. This is because we generically fill both of two bound states with pure imaginary eigenvalues when constructing the ground state. This is somewhat pathological in the sense that it contradicts with the Pauli's principle if one cares only about the real part of the energies. Nevertheless, because the imaginary parts of the two bound states differ, these modes are allowed to be occupied simultaneously, resulting in the anomalous behaviour of the dot density.

\begin{figure}[h]
\includegraphics[width=.5\textwidth ]{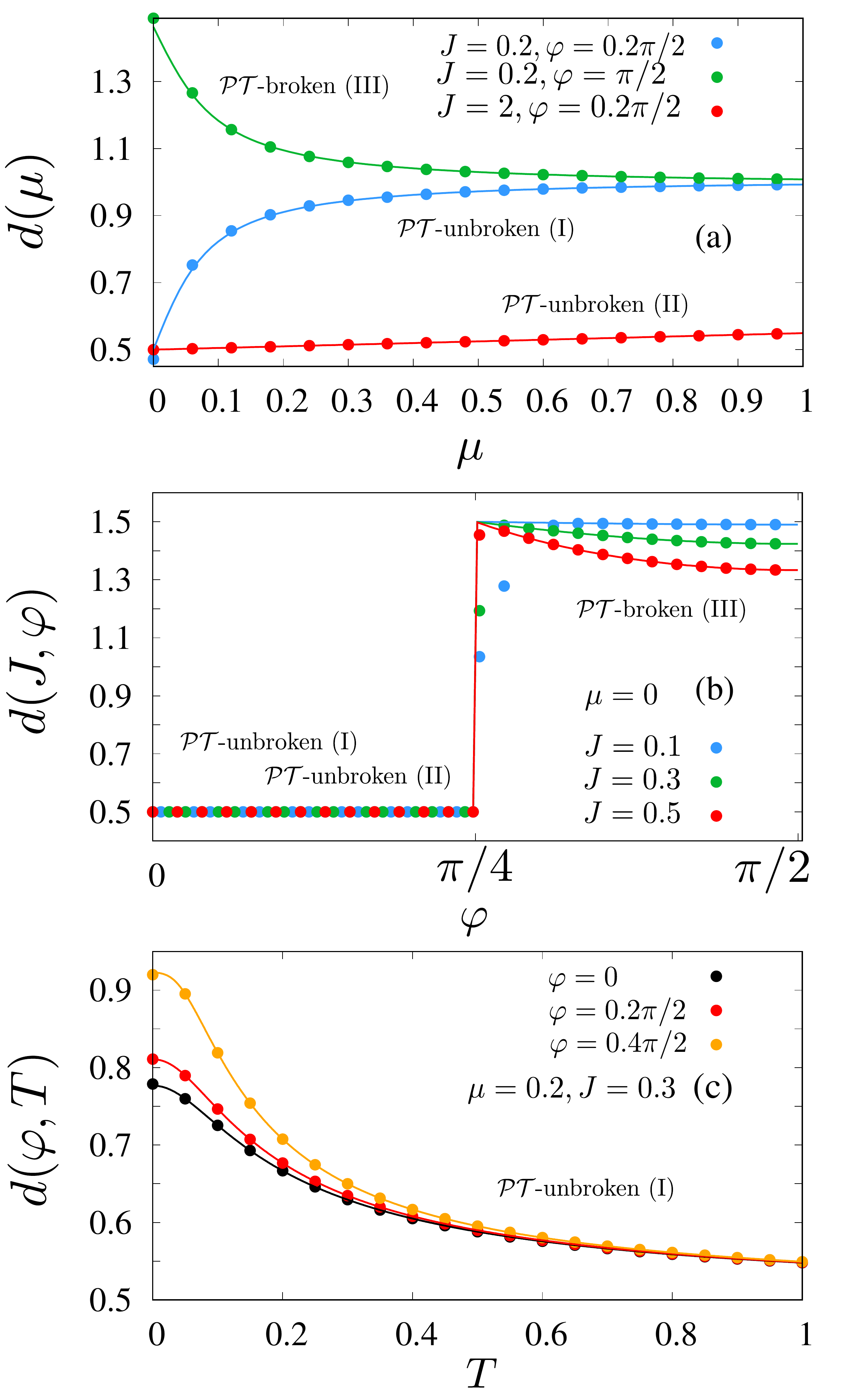}
\caption{Panel (a): dot density as a function of the chemical potential for different values of the coupling magnitude $J$ and argument $\varphi$ ($\gamma = J e^{i \varphi}$). Panel (b): dot density as a function of the coupling argument $\varphi$ for different values of the coupling magnitude $J$ and the fixed chemical potential $\mu=0$. A non-analytic behaviour appears as a jump between the $\mathcal{PT}$-unbroken and the $\mathcal{PT}$-broken phases in the vicinity of the line of exceptional points  $\varphi=\pi/4$. Numerical predictions depart from analytic curves near $\varphi=\pi/4$, but slowly converges to it with increasing a system size $N$, here $N=2019$. Panel (c): the dot density as a function of temperature $T$ for $\mu=0.2$ and $J=0.3$ , the color encodes fractions of $\varphi$ between real line $\varphi=0$ and the critical line $\varphi=\pi/4$.}
\label{dotdensity1}
\end{figure}

We also remark that the presence of bound states gives rise to another unusual phenomenon: the dot density no longer equals 1/2 at half-filling, indicating that particle-hole symmetry is 
{\it spontaneously broken}.  
This observation further implies a peculiar behavior of the magnetization $m=1/2-d$ . For the sake of simplicity, let us focus on the scaling limit $\gamma_1,\gamma_2,\mu\ll1$. When $\gamma_2$ crosses the value of $\gamma_1$ from below, the dot density jumps from $1/2$ to $3/2$, see Fig. \ref{dotdensity1}. This implies that the magnetization jumps from $0$ to $-1$. This is, in fact, similar to what is observed in \cite{ptkondo} where a jump of magnetisation associated with the $\mathcal{PT}$ phase transition was also found in the $\mathcal{PT}$ symmetric Kondo model. It is therefore natural to expect that both jumps are caused by the same mechanism, namely the breakdown of the particle-hole symmetry triggered by the $\mathcal{PT}$ phase transition. The emergence of edge modes with pure imaginary eigenvalues, which is associated to spontaneous broken particle-hole symmetry, was also previously observed in photonic graphene with gain and loss \cite{PhysRevA.98.042104}.

Having pointed out the peculiar behaviour of the dot density in the phase III, let us address a natural question that arises from the above observation: what is actually happening when $\mathcal{PT}$ symmetry is broken, and in particular, does Kondo screening disappear in the phase III or not? To answer this question, let us recall the phenomenology of Kondo screening in the phase I; Kondo screening in general refers to the fact that the impurity is healed with the leads in the IR, forming a spin singlet. In the scaling limit, this is typically signalled by the behaviour of the dot density in the IR, i.e. if the dot density goes to $1/2$ in the IR, we interpret it as the evidence that Kondo screening is taking place. This is in contrast to what happens in the UV, i.e. $d\to1$, which indicates that the spin is effectively free. Now, when $\mathcal{PT}$ symmetry is spontaneously broken, the IR and UV behaviours of the dot density largely differ from those in the phase I. Namely, along the flow from the UV to the IR, the value of the dot density changes from $1$ to $3/2$. We then {\it interpret} from this that the Kondo screening ceases to exist in the phase III. We, however, emphasise that this is not exactly what we expect from the previous studies on ISG. For instance, in  \cite{doi:10.1142/S0217751X93002265}, it was discovered that the starting and end points of the RG flow in ISG are characterised by almost the same fixed points, the $c=1$ CFT, with only difference being the compactification radii of them. Translating this situation into PTRLM, we are led to expect that the boundary RG flow in the phase III should start from the Neumann boundary condition (impurity being disconnected from the leads) and return to the same boundary condition again in the IR. This will be then reflected to the behaviour of $d$, namely $d$ should approach to $1$ in the IR, which is not quite what we observe. For now we have no convincing explanation to reconcile this dichotomy; one possibility is that the IR behaviour in the phase III is a peculiarity in the free theory, and the expected IR physics (i.e. return to the Neumann boundary condition) might be recovered by the inclusion of interaction.

As a final remark for the dot density, it is also worth mentioning what would happen to the dot density in the phase III if we were to fill just one of the bound states when constructing the ground state. It turns out that the value of the dot density is the same regardless of which bound state is filled, and, rather curiously, $d=1/2$ at half-filling. This implies that particle-hole symmetry is not broken in the phase III with this choice of the ground state, or at least, the breakdown of it, if any, cannot be deduced from the dot density at half-filling.

Another quantity that characterises the equilibrium property of the impurity system is the dot susceptibility defined by $\chi(T)=\left.\partial d/\partial\mu\right|_{\mu=0}$. From \eqref{dot2} and \eqref{dot3}, it can be easily seen that indeed the susceptibility follows Curie's law $\chi(T)\sim 1/T$, indicating that the impurity behaves like a free spin when temperature is high. Interesting physics emerges at low temperature; in any phase, we can readily see that the susceptibility reaches to the stationary value in the ground state $\chi(T=0)=1/(2\pi\tgamma)$, which is a hallmark of the Kondo screening. Therefore the existence of the $\mathcal{PT}$ phase transition cannot actually be inferred from the behaviour of the susceptibility. Notice also that the susceptibility blows up as the system approaches to the $\mathcal{PT}$ critical line defined by $\gamma_1=\gamma_2$. This is a genuine non-Hermitian effect, caused by the competition between the gain and loss terms.

\subsection{UV and IR expansions of the dot density at $T=0$}
At $T=0$, it is in fact possible to carry out the integrations explicitly (see Appendix \ref{dotdensity} for the details). For instance, In the phase I, we have the following UV expansion (small coupling)
\begin{align}\label{dphase1uv}
    d&=1+\frac{\Tilde{\Gamma}}{\pi(1-2\Tilde{\Gamma})}\arctan \frac{\sqrt{4-\mu^2}}{\mu}\n
    &\quad+\frac{1-\Tilde{\Gamma}}{\pi u(\Tilde{\Gamma})(1-2\Tilde{\Gamma})}\sum_{n=1}^\infty\frac{(-1)^n}{2n-1}u(\Tilde{\Gamma})^{2n}
\end{align}
where $u(\Tilde{\Gamma})=-\Tilde{\Gamma}\tan k_\mathrm{F}/(1-\Tilde{\Gamma})=\Tilde{\Gamma}\sqrt{4-\mu^2}/(\mu(1-\Tilde{\Gamma}))$ which is a monotonically increasing function of $\Tilde{\Gamma}$ in the phase I (i.e. $\Tilde{\Gamma}<1$ ). Note that the convergence of radius of the series is $u(\Tilde{\Gamma})<1$. The IR expansion reads, on the other hand,
\begin{align}\label{dphase1ir}
   d&=\frac{1}{2}-\frac{\Tilde{\Gamma}}{\pi(1-2\Tilde{\Gamma})}\arctan \frac{\mu}{\sqrt{4-\mu^2}}\n
    &\quad-\frac{u(\Tilde{\Gamma})(1-\Tilde{\Gamma})}{\pi (1-2\Tilde{\Gamma})}\sum_{n=1}^\infty\frac{(-1)^n}{2n-1}u(\Tilde{\Gamma})^{-2n},
\end{align}
which converges only when $u(\Tilde{\Gamma})<1$.
Similarly, in the phase III, the UV and IR expansions are
\begin{align}\label{dphase3uv}
    d&=1+\frac{\Tilde{\Gamma}}{\pi(1-2\Tilde{\Gamma})}\arctan \frac{\sqrt{4-\mu^2}}{\mu}\n
    &\quad-\frac{1-\Tilde{\Gamma}}{\pi v(\Tilde{\Gamma})(1-2\Tilde{\Gamma})}\sum_{n=1}^\infty\frac{(-1)^n}{2n-1}v(\Tilde{\Gamma})^{2n}
\end{align}
and
\begin{align}\label{dphase3ir}
   d&=\frac{1}{2}+\frac{1-\Tilde{\Gamma}}{1-2\Tilde{\Gamma}}-\frac{\Tilde{\Gamma}}{\pi(1-2\Tilde{\Gamma})}\arctan \frac{\mu}{\sqrt{4-\mu^2}}\n
    &\quad+\frac{v(\Tilde{\Gamma})(1-\Tilde{\Gamma})}{\pi (1-2\Tilde{\Gamma})}\sum_{n=1}^\infty\frac{(-1)^n}{2n-1}v(\Tilde{\Gamma})^{-2n},
\end{align}
where $v(\Tilde{\Gamma})=\Tilde{\Gamma}\tan k_\mathrm{F}/(1-\Tilde{\Gamma})=-\Tilde{\Gamma}\sqrt{4-\mu^2}/(\mu(1-\Tilde{\Gamma}))$, and the UV and IR expansions are valid when $v(\Tilde{\Gamma})<1$ and $v(\Tilde{\Gamma})>1$, respectively. We note that, in the scaling limit $\gamma_1,\gamma_2,\mu\ll1$, \eqref{dphase1uv} and \eqref{dphase1ir} reproduce the known expression \cite{boulat1,camacho1} in the Toulouse limit of the anisotropic Kondo model. Similar formulae can also be obtained by the same manipulations in the phase II, and presented in Appendix \ref{dotdensity}. Profiles of the dot density at $T=0$ as a function of $\mu$ in different phases are depicted in the panel (a) of Fig. \ref{dotdensity1}.

\section{$g$-function and boundary free energy}
\subsection{$g$-function}
\begin{figure}[h]
\includegraphics[width=0.5\textwidth]{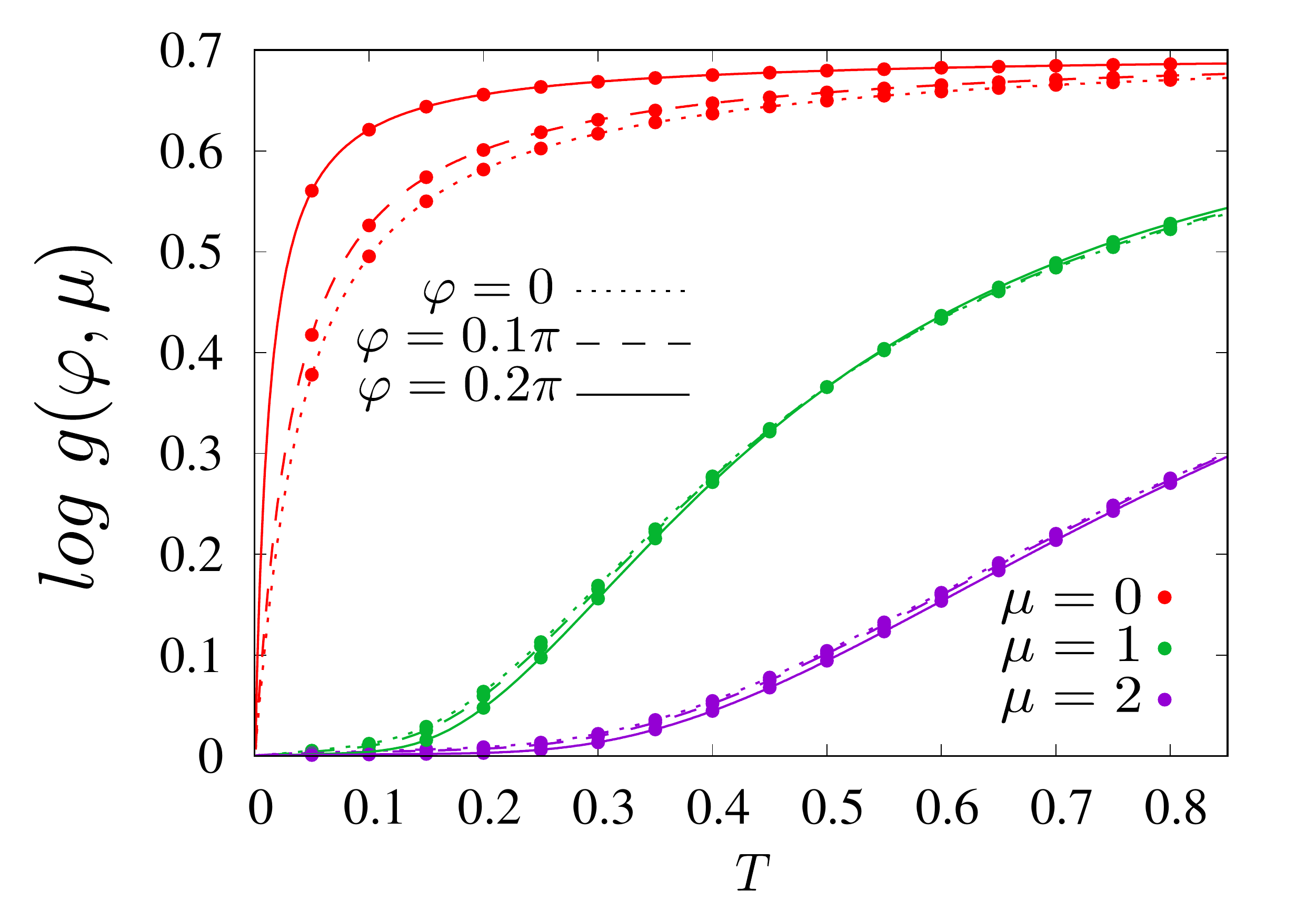}
\caption{Impurity entropy $log\ g $ as a function of temperature $T$ for different values of chemical potential $\mu$.  The color encodes  $\mu$, while the line types encodes value of a coupling constant argument $\varphi$ ($\gamma=J e^{i\varphi}$). Points represent numerics, while lines represent analytical result Eq.\eqref{gfunc1}.}
\label{fig:gfunc}
\end{figure}

It is always instructive to study the $g$-function to see the UV and IR behaviour of a given impurity system. It is known that, in quantum impurity systems in the scaling regime, the $g$-function undergoes a monotonic change under the boundary renormalisation group flow, satisfying $g_\mathrm{UV}<g_\mathrm{IR}$ \cite{affleck1}. In PTRLM, we can readily observe the crossover physics of the $g$-function as $T$ increases. The $g$-function is defined in terms of the impurity entropy
\begin{equation}\label{gfunc}
    \log g=S_{\rm imp}=S-S_0
\end{equation}
where $S$ is the thermodynamic entropy of the RLM, and $S_0$ is the same quantity without the impurity (i.e. two decoupled leads). Recall that the partition function is given by $Z=\prod_{\sigma,k_n}(1+e^{-\beta(E_k-\mu)})$, where $k_n$ satisfies
\begin{equation}
k_n=\frac{(2n+1)\pi}{N+2}+\frac{k\delta_\sigma(k)}{N+2},\quad n=1,\cdots,\frac{N+2}{2}
\end{equation}
Once again, we note that $k$ implicitly depends on $\sigma$. Then, the root density $\rho(k)$ is
\begin{align}
    \rho_\sigma(k_n)&=\lim_{N\to \infty}\frac{1}{(N+2)(k_{n+1}-k_n)} \n
    &=\frac{1}{2\pi}\Big(1+\frac{2}{N+2}\frac{1-\Gamma_\sigma^2}{\Delta_\sigma(k)}+\mathcal{O}(N^{-2})\Big).
\end{align}
Therefore, under the thermodynamic limit, the partition function becomes
\begin{align}
    \log Z&=(N+2)\sum_{\sigma,k_n}\frac{1}{N+2}\rho_\sigma(k_n)\log(1+e^{-\beta(E_{k_n}-\mu)}) \n
    &=2(N+2)\int_{0}^\pi\frac{\dd k}{2\pi}\log(1+e^{-\beta(E_{k}-\mu)}) \n
    &\quad + \sum_\sigma\int_{0}^\pi\frac{\dd k}{2\pi}\frac{1-\Gamma^2_\sigma}{\Delta_\sigma(k)}\log(1+e^{-\beta(E_{k}-\mu)}),
\end{align}
where the non-extensive part represents the impurity contribution. Since, without the impurity, $\log Z$ is simply $\log Z=(N+2)\int_0^\pi\frac{\dd k}{\pi}\log(1+e^{-\beta(E_{k}-\mu)})$, using the formula $S=\log Z-\beta \partial_\beta\log Z$, the $g$-function can be obtained as
\begin{align}\label{gfunc1}
    \log g&=-2\int_0^\pi\frac{\dd k}{\pi}\Big[\frac{(2\tgamma-1)\cos2k-1}{4(\tgamma^2+(1-2\tgamma)\cos^2k)}+\frac{1}{2}\Big] \n
    &\quad\times\Bigg[\log(1+e^{-\beta(E_{k}-\mu)})+\frac{\beta(E_k-\mu)}{1+e^{\beta(E_k-\mu)}}\Bigg].
\end{align}
One can readily confirm that when $\beta\gg 1$, $\log g\to0$, indicating that at low-temperature, the impurity is effectively hybridised with the wires. On the other hand, for $\beta\ll1$, $\log g\to \log2$, which suggests that at high-temperature, the impurity is decoupled from the wires, attaining two degrees of freedom (occupied or unoccupied dot state), see Fig. \ref{fig:gfunc}.

Next we treat the phase II. The additional term $S_{{\rm imp},{\rm b}}$ to \eqref{gfunc} comes from two bound states, reading
\begin{align}\label{gfunc2}
    S_{{\rm imp},{\rm b}}&=\log[1+e^{2\beta\mu}+2e^{\beta\mu}\cosh(\beta E_{\rm b})] \n
    &\quad-2\beta e^{\beta\mu}\frac{\mu e^{\beta\mu}+\mu\cosh\beta E_{\rm b}+E_\mathrm{b}\sinh\beta E_{\rm b}}{1+e^{2\beta\mu}+2e^{\beta\mu}\cosh(\beta E_{\rm b})}.
\end{align}
Therefore the $g$-function in the phase II is the sum of \eqref{gfunc1} and \eqref{gfunc2}
\begin{align}\label{gfunc3}
    \log g&=-2\int_0^\pi\frac{\dd k}{\pi}\Big[\frac{(2\tgamma-1)\cos2k-1}{4(\tgamma^2+(1-2\tgamma)\cos^2k)}+\frac{1}{2}\Big] \n
    &\quad\times\Bigg[\log(1+e^{-\beta(E_{k}-\mu)})+\frac{\beta(E_k-\mu)}{1+e^{\beta(E_k-\mu)}}\Bigg]+S_{{\rm imp},{\rm b}}
\end{align}
The asymptotic behaviour in the phase II turns out to be the same as in the phase I. This is because the bound state energy satisfies $E_{\rm b}>2$ in the phase II ($\gamma^2_1-\gamma^2_2>1$), and as such $E_{\rm b}>\mu$ always holds when the ground state is nontrivial (the ground state becomes nontrivial only when $\mu<E_k$ for some $k\in[-\pi,\pi]$, i.e. $0<\mu<2$.).
\subsection{Boundary free energy}
Now, we turn our attention to the phase III. As argued in the previous section, we do not have any invariant statistical measure when $\mathcal{PT}$-symmetry is spontaneously broken, and hence, $g$-function is not suitable for analysing the UV and IR behaviours of the system. That being said, there exists an alternative quantity, the boundary free energy, from which we can infer the two asymptotic behaviours of the system. The boundary free energy $f$ can be obtained from the definition $m=\frac{\partial f}{\partial \mu}$ (note that $\mu$ in the RLM language is the opposite of the natural magnetic field in the corresponding  (anisotropic) Kondo problem): we will compare the behaviours of the boundary free energy in the phase I and III in the rest of this section. To avoid unnecessary complication, let us focus on the real coupling case $\gamma=\gamma_1$ (phase I) and the pure imaginary coupling case $\gamma=\ii\gamma_2$ (phase III) in the scaling limit. Then, according to \eqref{dphase1uv} and \eqref{dphase3uv}, the boundary free energy in the phase I and III are given by, respectively
\begin{align}
    f_\mathrm{I}&=-\frac{\mu}{2}+\frac{2\gamma^2_1}{\pi}\Big(\frac{\mu}{2\gamma^2_1}\arctan{\frac{2\gamma^2_1}{\mu}}+\frac{1}{2}\log\Big(1+\frac{\mu^2}{4\gamma^4_1}\Big)\Big)\n
    f_\mathrm{III}&=-\frac{\mu}{2}-\frac{2\gamma^2_2}{\pi}\Big(\frac{\mu}{2\gamma^2_2}\arctan{\frac{2\gamma^2_2}{\mu}}+\frac{1}{2}\log\Big(1+\frac{\mu^2}{4\gamma^4_2}\Big)\Big).
\end{align}
Accordingly their asymptotic behaviors are
\begin{align}
    f_\mathrm{I}&\to\begin{dcases*}
    -\frac{\mu}{2} & $\gamma_1\to0$\\
   0 & $\gamma_1\to\infty$
    \end{dcases*}\label{f1}\\
    f_\mathrm{III}&\to\begin{dcases*}
    -\frac{\mu}{2} & $\gamma_2\to0$\\
   -\mu & $\gamma_2\to\infty$\label{f2}
    \end{dcases*}.
\end{align}
Recall that the boundary free energy $f$ encodes information on the impurity degrees of freedom at the UV (resp. IR) fixed point when $\gamma\to0$ (resp. $\gamma\to\infty$). In the spin $j$ anisotropic Kondo model, it is known that $f$ behaves as $f\to-j\mu$ and $f\to-(j-\frac{1}{2})\frac{\mu}{g}$ near the UV and IR fixed points, respectively. Note that near the IR fixed point, the spin is renormalised by the Luttinger coupling constant $g$, which comes from the fact that the underlying model is anisotropic, i.e. $SU(2)$ symmetry is spoiled.

The choice $j=\frac{1}{2}$, $g={1\over 2}$ corresponds to the scaling limit of the resonant level model for which we recover \eqref{f1}. This is essentially what we saw in \eqref{gfunc1}: near the UV fixed point, the impurity spin is effectively decoupled while near the IR fixed point, the impurity is screened and accordingly $f\to0$. The UV behaviour of $f$ in the phase III is same as that in the phase I, as the impurity is decoupled anyways even if coupling constants are complex. The IR behaviour, however, comes as a surprise; the impurity remains active even when the couplings are strong enough. This is in accordance with the anomalous behaviour of the dot density, and again we understand this as the signal of the lack of Kondo screening. Remarkably, we also observe \cite{future} that this peculiar IR behaviour of $f$ persists even away from the free limit $g=\frac{1}{2}$ and for arbitrary spin $j$ in the anisotropic Kondo model, with $f\to-j\frac{\mu}{g}$, suggesting that this is a universal phenomenon in quantum impurity systems when $\mathcal{PT}$ symmetry is broken.

\section{Loschmidt echo in the $\mathcal{PT}$-symmetric RLM}
One of the simplest objects that encodes nontrivial information on dynamics is the {\it Loschmidt echo}, which is defined by
\begin{equation}\label{Echo}
    \overline{{\mathcal{L}}}(t)={}_0\langle \mathrm{GS}|e^{\ii tH_0}|\Psi(t)\rangle,
\end{equation}
where $|GS\rangle_0$ denotes the ground state of two disconnected chains $H_0=H_\mathrm{L}+H_\mathrm{R}$. Here $|\Psi(t)\rangle$ denotes the normalized time-evolved state $|\Psi(t)\rangle=|\psi(t)\rangle/|\langle\Tilde{\psi}(t)|\psi(t)\rangle|$, where $|\psi(t)\rangle=e^{-\ii tH}|\mathrm{GS}\rangle_0$, and $\langle\Tilde{\psi}(t)|$ is its associated state \eqref{associated}. Note that the norm is 1 as long as $\mathcal{PT}$ symmetry is not broken, in which case dynamics is unitary. Practically the norm plays a role of taming the exponential growth of the Loschmidt echo in the phase III. We note that a similar normalization is necessary whenever Loschmidt echos for Hermitian hamiltonians are calculated in imaginary time using CFT \cite{St_phan_2011} or integrability methods \cite{PIROLI2018454}: in that case, phase terms $e^{iE_0t}$ indeed cannot be neglected as they acquire exponential growth. Loschmidt echo with this protocol was also studied previously in noninteracting impurity systems using the form factor approach \cite{PhysRevLett.110.240601} and the finite volume free fermion technique which is akin to ours \cite{2019arXiv191101926G}.
 
The Loschmidt echo simply characterises the return probability of the initial state. Each chain can be diagonalised easily as
\begin{equation}
    H_\mathrm{L}=\sum_{k>0}E_kd^\dagger_{1k}d_{1k},\quad H_\mathrm{R}=\sum_{k>0}E_kd^\dagger_{2k}d_{2k},
\end{equation}
where canocical femionic operators are given by
\begin{equation}
    d_{1k}=\sum_{x>0}\phi(k;x) \,c_x,\quad d_{2k}=\sum_{x<0}\phi(k;x) \,c_x
\end{equation}
with $\phi(k;x)=2\sin kx/\sqrt{N+2}$ and $k$ is quantised as $k_n=2\pi n/(N+2), n=1,\dots,(N+2)/2$. Therefore $H_0$ can be written as $H_0=\sum_{k>0}E_k(d^\dagger_{1k}d_{1k}+d^\dagger_{2k}d_{2k})$ with the ground state
\begin{equation}
    |\mathrm{GS}\rangle_0=\prod_{n\in\mathrm{FS}}c^\dagger_{k_n}, \quad c_{k_n}=\begin{cases}
    d_{1k_n} & n>0\\
    d_{2k_n} & n<0
    \end{cases},
\end{equation}
where the integer set $\mathrm{FS}=\{-M,\cdots,-1,1,\cdots,M\}$ corresponds to the modes comprising Fermi seas of each lead with $M=(N+2)/2$. Invoking that the Loschmidt echo \eqref{Echo} has a structure of the Slater determinant, we immediately realise that $\mathcal{L}(t)$ can be written as a determinant of a matrix involving overlap matrices
\begin{equation}
    \mathcal{L}(t)=e^{\ii tE_\mathrm{GS}}\det_{i,j\in\mathrm{FS}}\mathtt{M}(k_i,k_j;t),
\end{equation}
where $\mathtt{M}(k_i,k_j;t)=\sum_{m}e^{-\ii E_{m}t}\mathtt{L}^*(k_i;m)\mathtt{R}(k_{j};m)$. Here the overlap matrices are defined as $\mathtt{L}^*(k;m)=\sum_x\vartheta(kx)\phi(|k|;x)L^*(m;x)$ and $\mathtt{R}(k;m)=\sum_x\vartheta(kx)\phi(|k|;x)R(m;x)$, where $\vartheta(x)$ is the Heaviside step function. After carefully evaluating the matrix $\mathtt{M}(k_i,k_j;t)$ (see Appendix \ref{echosec} for the details), we end up with the following Fredholm determinant
\begin{equation}\label{echophase1}
     \overline{\mathcal{L}}(t)=\begin{dcases*}
     \det_{i,j\in\mathrm{FS}}\left(1+\frac{2\pi}{N+2}\mathtt{B}(t)\right)_{ij} & Phase I \\ 
     \det_{i,j\in\mathrm{FS}}\left(1+\frac{2\pi}{N+2}\mathtt{B}(t)+\mathtt{B}^\mathrm{re}_\mathrm{b}(t)\right)_{ij} & Phase II\\
     \frac{ \det_{i,j\in\mathrm{FS}}\left(1+\frac{2\pi}{N+2}\mathtt{B}(t)+\mathtt{B}^\mathrm{im}_\mathrm{b}(t)\right)_{ij}}{\sqrt{|\det_{i,j\in\mathrm{FS}}\left(1+\frac{2\pi}{N+2}\Tilde{\mathtt{N}}(t)\right)|}} & Phase III\\
     \end{dcases*}
\end{equation}
where the matrix $\mathtt{B}(t),\mathtt{B}^\mathrm{re}_\mathrm{b}(t),\mathtt{B}^\mathrm{im}_\mathrm{b}(t)$, and $\Tilde{\mathtt{N}}(t)$ are defined in \eqref{amatrix}, \eqref{bre}, \eqref{bim}, and \eqref{binorm} respectively. We are primarily interested in the behaviour of $\sqrt{t}|\overline{\mathcal{L}}(t)|^2$ as a function of $t$. The reason for having the additional factor $\sqrt{t}$ is the following: the large $t$ behaviour of the echo is dominated by the low-energy process, under which case the dot is fully healed with the leads. As such, for large $t$, $\overline{\mathcal{L}}(t)$ is expected to match with that describing the quench from the two disconnected chains to the fully homogeneous chain, which, for large enough $t$, goes as $t^{-1/4}$ \cite{PhysRevLett.110.240601}, where the exponent is related to the scaling dimension of the boundary condition changing operator from free to fixed boundary condition in the critical boundary Ising model \cite{PhysRevLett.110.240601}. We therefore multiply $\sqrt{t}$ to $|\overline{\mathcal{L}}(t)|^2$ in order to cancel out such decay.

\begin{figure}
\includegraphics[width=0.5\textwidth]{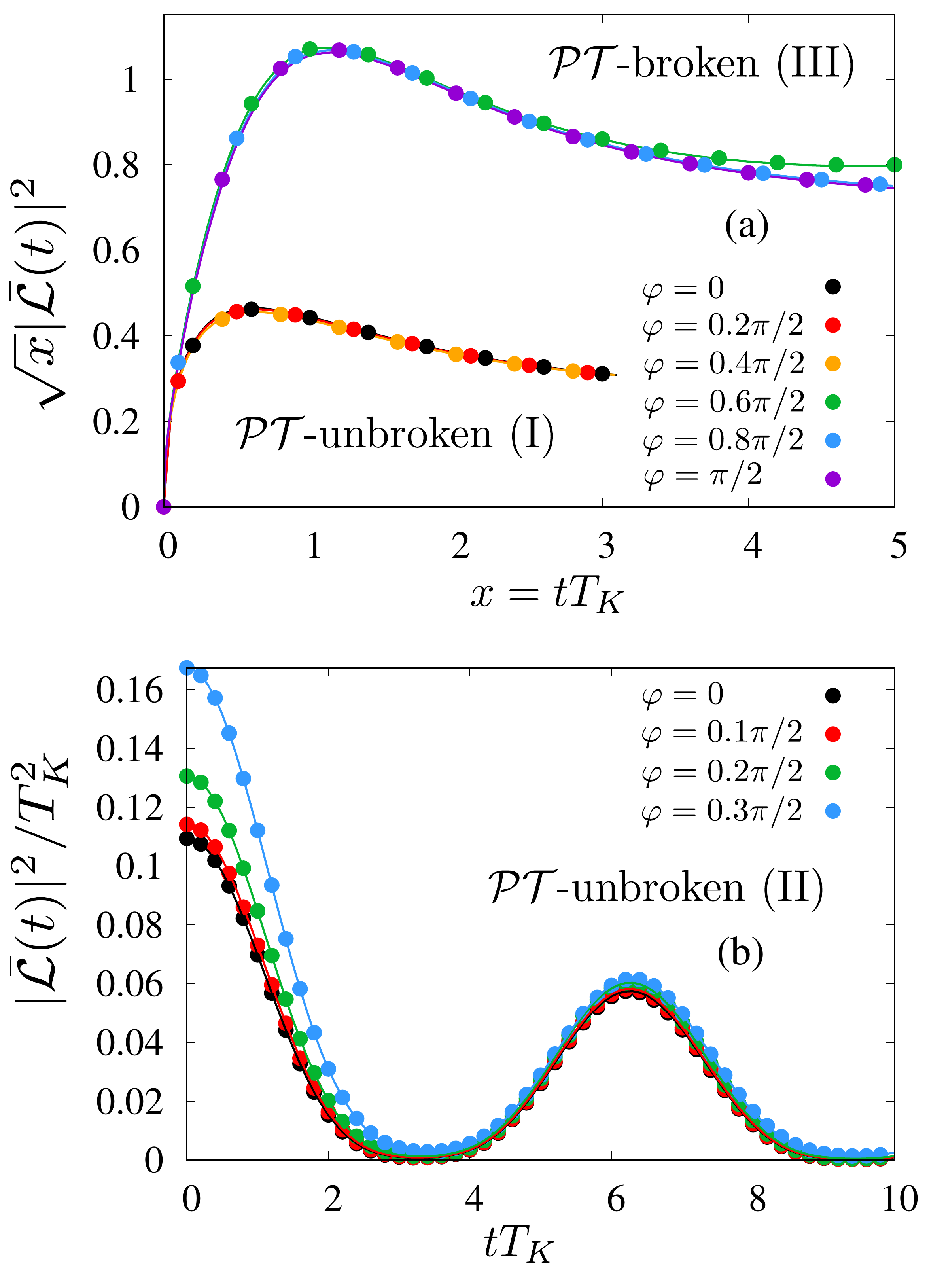}
\caption{Panel (a): rescaled Loschmidt echo $\sqrt{x}|\bar{\mathcal{L}}(x)|^2$ as a function of rescaled time $x=tT_K$ in phases I and III for different values of $\varphi$  (coupling constant $\gamma=0.2e^{\ii\pi \varphi}$). \\
Panel (b): rescaled Loschmidt echo $|\bar{\mathcal{L}}(x)|^2 / T_K^2$ (scaling is  $1/T_K^2$) as a function of rescaled time in phase II ($\gamma=2 e^{i \varphi}$). Scaling factor is chosen to be $T_K = |E_b|$.
The colors label the
values of $\varphi$, lines represent Fredholm determinant Eq.\eqref{echophase1}, while dots obtained by numerics.}
\label{fig:echo1}
\end{figure}

Let us first look at the behaviour in the phase I. We find  that the echo exhibits a crossover behaviour whose time scale is controlled by the generalised Kondo temperature $T_\mathrm{K}$. The temperature is in fact nothing but the absolute value of the eigenenergies of the bound states: $T_\mathrm{K}=|E_\mathrm{b}|$, appearing in the phase II and III. This phenomenon, which can be regarded as a dynamical manifestation of the Kondo screening, was already observed in \cite{PhysRevLett.110.240601,PhysRevB.90.115101}, but a new observation here is that such dynamical Kondo effect persists even away from the Hermitian case, as long as the $\mathcal{PT}$ symmetry of the system is unbroken. Furthermore, we also observe that the echo possesses the universal scaling form $\sqrt{tT_\mathrm{K}}f(tT_\mathrm{K})$, which was also shown in \cite{PhysRevLett.110.240601} in the Hermitian case. We depict the behaviour of the echo in the phase I in \ref{fig:echo1} where we compare the analytic results \eqref{echophase1} with independent free fermion numerics. We not only observe a beautiful agreement but also clearly see that the echo follows the universal curve, which has a crossover around $t\sim 1/T_\mathrm{K}$. Such a crossover is a simple consequence of interpolating the UV and IR behaviour of the echo: for short time (UV), the dot is effectively screened from the chains, hence the system behaves as two disconnected chains, giving rise to a sharp build-up of the echo. For large time (IR), however, the echo decays as $t^{-1/2}$ for the reason above, thus  presents a bump when undergoing the transition from the UV to IR time scale. It is also noteworthy that, since $E_\mathrm{b}\sim \tgamma$, the time needed for the crossover can be arbitrarily long near the $\mathcal{PT}$ critical line $\gamma_1=\gamma_2$. 

We next turn to the phase II. In this phase the Loschmidt echo displays persistent oscillations whose frequency is controlled by the Kondo temperature $T_K$. These oscillations are caused by the bound states, thereby spoiling the dynamical Kondo screening which was observed in the phase I. With the appropriate rescaling of the Loschmidt echo, which can be found numerically, we observe that the curves for different $\varphi$ can collapse onto a single universal curve for large $t$ (see the panel (b) in Fig.\ref{fig:echo1}).

Having analysed the Loschmidt echo in the phase I and II, let us finally deal with the most controversial phase: the phase III. In this regime the echo is dominated by the contribution from the eigenmode with positive imaginary energy, hence $\mathcal{L}(t)$ exponentially grows. The norm then serves to cancel out such explosion, yielding a finite result. Curiously, the normalised echo seems to show a similar crossover physics as time increases, and rather surprisingly, it turns out to decay in the same fashion as in the phase I, i.e. decays as $\overline{\mathcal{L}}(t)\sim t^{-1/4}$. Such a crossover behaviour in the $\mathcal{PT}$-broken regime can be expected by invoking the similar argument as in the $\mathcal{PT}$-unbroken regime. What is surprising here however is that, with the division by the nontrivial norm, the asymptotic behaviour of the normalised echo follows the same pattern as in the phase I, and the position of the bump is still controlled by the absolute value of the Kondo temperature. We emphasise that whether the mechanism of this decay is the same as in the phase I is not at all obvious, and necessitates a further understanding on the phase III. The collapse of curves onto the single universal curve also occurs in this phase in the scaling regime, see Fig. \ref{fig:echo1}. 

\section{Approaching the different phases  via analytic continuation}\label{anacont}

First, we note that in all regimes, the dot density is an analytical function of $\tilde{\Gamma}$ at fixed $\mu$. $d$ admits two different expansions depending on whether the coupling $\tilde{\Gamma}$ is small (UV) or large (IR), but these are two expansions of the same function, whose singularities lie in general elsewhere (more on this below). For instance, we have in regime I
\begin{eqnarray}
d=1+{\tilde{\Gamma}\over \pi(1-2\tilde{\Gamma})} \arctan {\sqrt{4-\mu^2}\over \mu}\nonumber\\-{1-\tilde{\Gamma}\over \pi(1-2\tilde{\Gamma})}\arctan u(\tilde{\Gamma})~~\mathrm{UV}\label{UV-I}\\
d=1+{\tilde{\Gamma}\over \pi(1-2\tilde{\Gamma})} \arctan {\sqrt{4-\mu^2}\over \mu}\nonumber\\-{1-\tilde{\Gamma}\over \pi(1-2\tilde{\Gamma})}\left({\pi\over 2}-\arctan {1\over u(\tilde{\Gamma})}\right)~~\mathrm{IR}\label{IR-I}
\end{eqnarray}
where recall $u(\tilde{\Gamma})={\tilde{\Gamma}\sqrt{4-\mu^2}\over \mu(1-\tilde{\Gamma})}$. We see that the IR expansion is obtained from the UV one by using the continuation of the function $\arctan$ on the positive real axis when its argument gets larger than one. 

The same feature holds in regimes II and III. For instance in regime III we now have 
\begin{eqnarray}
d=1+{\tilde{\Gamma}\over \pi(1-2\tilde{\Gamma})} \arctan {\sqrt{4-\mu^2}\over \mu}\nonumber\\+{1-\tilde{\Gamma}\over\pi(1-2\tilde{\Gamma})}\arctan v(\tilde{\Gamma})~~\mathrm{UV}\label{UV-III}\\
d=1+{\tilde{\Gamma}\over \pi(1-2\tilde{\Gamma})} \arctan {\sqrt{4-\mu^2}\over \mu}\nonumber\\-{1-\tilde{\Gamma}
\over \pi(1-2\tilde{\Gamma})}\left(-{\pi\over 2}+\arctan {1\over v(\tilde{\Gamma})}\right)~~\mathrm{IR}\label{IR-III}
\end{eqnarray}
where $v(\tilde{\Gamma})=-\tilde{\Gamma}\sqrt{4-\mu^2}/(\mu(1-\tilde{\Gamma}))=-u(\tilde{\Gamma})$. Here again we see that the IR expansion is obtained by ``straightforward'' continuation of the UV one.

This feature is common to other problems of this type in the Hermitian case: it has been observed in the Kondo and  anisotropic Kondo models, and the problem of tunneling between edges in the fractional quantum Hall effect, at least in the scaling limit (where, in particular, the FQHE tunneling problem is described by the boundary sine-Gordon model \cite{PhysRevLett.74.3005}). It is expected whenever the short coupling expansion of a physical quantity admits a finite radius of convergence where there is however no physical singularity.

More surprising maybe is the fact that the UV expansion of $d$ in regimes II and III can be obtained as well by analytical continuation of the UV expansion in regime I. This is clear if one considers for instance the UV expansions in regime I and III : the result in equation (\ref{UV-III}) is in fact the same as  the one in (\ref{UV-I})  when expressed in terms of $\tilde{\Gamma}$. The result is in fact the same in regime II as well, and completely independent of the phase of $\Gamma$. This result is in fact quite reasonable if one imagines calculating a quantity such as $d$ perturbatively in $\tilde{\Gamma}$: this will produce a series integer powers of $\tilde{\Gamma}$, with amplitudes given by integrals of various correlations functions over the vacuum at $\tilde{\Gamma}=0$. Since this vacuum does not depend on $\tilde{\Gamma}$, there should be in fact a unique UV expansion for $d$ (and other such perturbative physical quantities). 

On the other hand, one has to be very careful with what happens in the IR regime. To obtain the  results at large coupling in regime III from those in regime I, one needs to pay attention to the determination of the $\arctan$ involved in the problem. Specifically, starting with equation (\ref{IR-I}), the corresponding result in   regime III is obtained not only by observing that $u(\tilde{\Gamma})=-v(\tilde{\Gamma})$: instead one must set 
\begin{equation}
\arctan {1\over u(\tilde{\Gamma})}\to \pi-\arctan{1\over v(\tilde{\Gamma})}\label{extra}
\end{equation}
even though we recall that $v(\tilde{\Gamma})=-u(\tilde{\Gamma})$. What happens of course is that the $\arctan$ function admits multiple determinations, and to see what happens to analytical continuations requires following trajectories on the corresponding Riemann surface. In other words, sending $z$ into $e^{i\pi} z$ and $z$ into $1/z$ are two operations that do not commute for the functions $\arctan z={1\over 2i}\ln{1+iz\over 1-iz}$, whose Riemann surface  is branched along the imaginary axis for $|z|\geq1$. 

Extra factors of $\pi$ such as in (\ref{extra}) will be needed only after $\hbox{arg}~ u(\tilde{\Gamma})>{\pi\over 2}$. In our model where $u$ is real\footnote{There are variants of the model which are not PT symmetric, such as the RLM with a complex coupling $\gamma$. In this case, $d$ can be expanded in powers of a variable similar to $u$, and proportional to $\gamma^2$. For $\gamma =Je^{i\varphi}$, the transition between the equivalent of regime I and regime III occurs when $\hbox{arg}\gamma^2={\pi\over 2}$, i.e. $\varphi={\pi\over 4}$. This will be discussed elsewhere.}, this corresponds to  $\varphi>{\pi\over 2}$, i.e. regime III indeed.

We can predict from this that the results in regime II at large coupling are obtained by the same formulas as those in regime I not only in the UV (as argued at the beginning) but also in the IR - this is detailed in the appendix.

\begin{figure}
\includegraphics[width=0.5\textwidth]{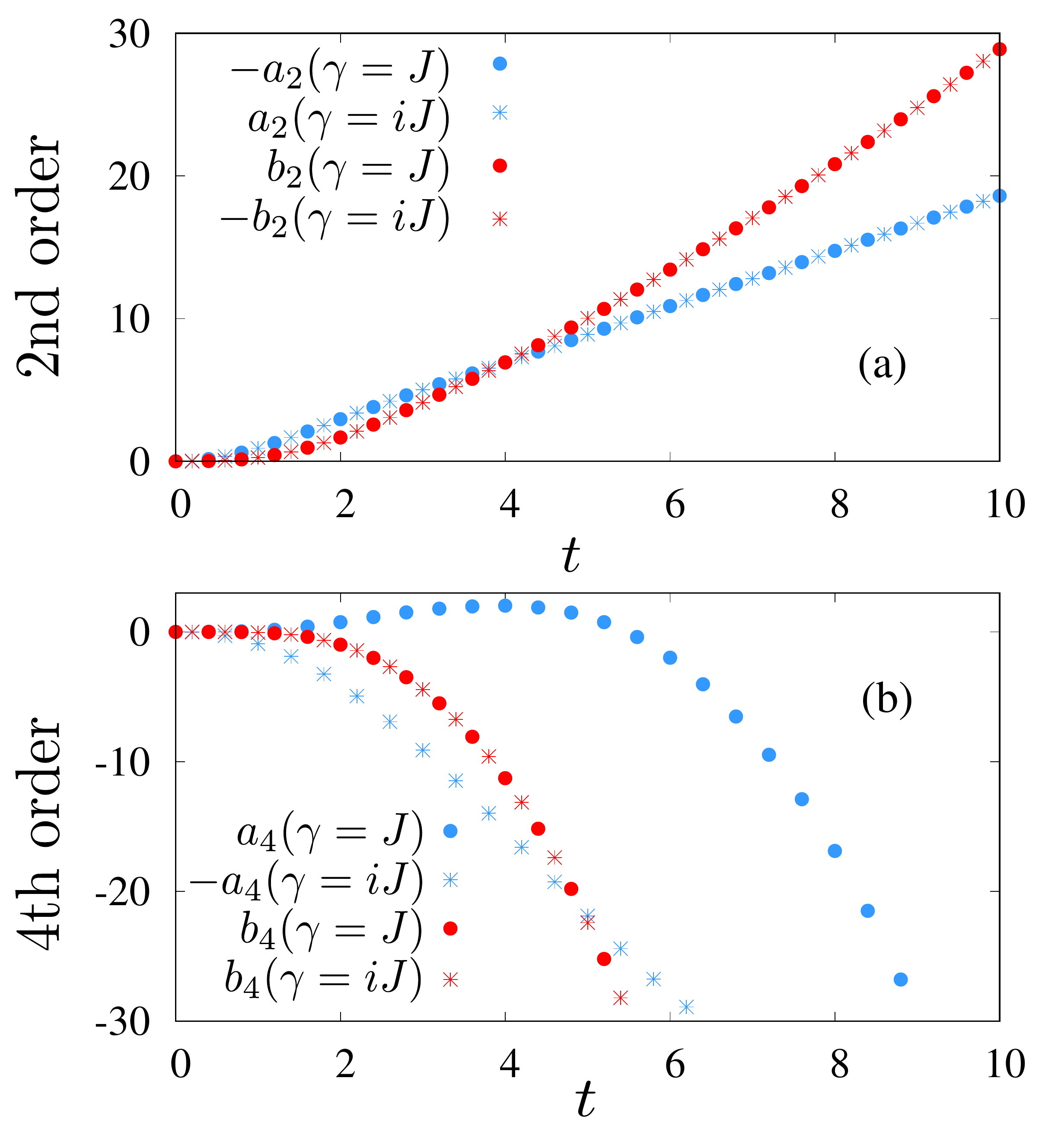}
\caption{Taylor coefficient $a_2, b_2$ and $a_4, b_4$ as a function of time for phase I $(\text{real coupling } \gamma = J)$ and phase III $(\text{imaginary coupling }\gamma = i J)$. Lower order coefficients are in agreement with analytical continuation picture (panel a), whereas the 4-th order coefficient $a_4$ behaves differently for real and imaginary couplings.}
\label{fig:coeff}
\end{figure}

Finally, contrary to what we expect, we find that analytic continuation does not seem to work for out-of-equilibrium quantities. To test the idea, we numerically expand two Loschmidt echoes and compare coefficients of Taylor series, one with $\gamma\in\mathbb{R}$ (Hermitian, phase I) and another one with $\gamma\in\ii\mathbb{R}$ (pure imaginary coupling, phase III) in terms of $\gamma$. We expect that, for analytic continuation to work, the expansion of the latter case be obtained by performing a transformation $\gamma=J\mapsto\ii J$ to the former one. We study the real and imaginary parts of the Loschmidt echo $\bar{\mathcal{L}}$ separately, each of which is given by
\begin{eqnarray}
\mathcal{R}e\bar{\mathcal{L}}(t)=a_0 + a_1\gamma + a_2\gamma^2 + a_3 \gamma^3 + a_4 \gamma^4 + \mathcal{O}(\gamma^5) \quad \nonumber \\ 
\mathcal{I}m\bar{\mathcal{L}}(t)=b_0 + b_1\gamma + b_2\gamma^2 + b_3 \gamma^3  + b_4 \gamma^4 + \mathcal{O}(\gamma^5). \quad 
\end{eqnarray}
The coefficient $a_0 = 1$ corresponds to the case $\gamma =0$, i.e. the initial and the time evolved states are equal, so $\la \psi_0 | \psi(t)\ra = 1$. 
 Note also that coefficients corresponding to the odd powers of $\gamma$ are zeroes $a_1 = a_3 = b_0 = b_1 = b_3 = 0$, indicating that the Loschmidt echo is a function of $\tilde{\Gamma}$ in both the phase I and III at least in the scaling regime. Nontrivial coefficients appear with the even powers of $\gamma$, and presented in
Fig.\ref{fig:coeff}. We first notice that $a_2(\gamma=J)=-a_2(\gamma=\ii J)$ and $b_2(\gamma=J)=-b_2(\gamma=\ii J)$ always hold at any time, which indicates that the analytic continuation works at this order. The next order, the coefficients of  $\gamma^4$, however, clearly invalidates the continuation. Namely, in Fig \ref{fig:coeff} panel (b), we observe a firm agreement for imaginary parts of the Loschmidt echo in the phases I and III, while the curves representing the real parts start departing as soon as $t>0$. For now we have no explanation as to why the continuation works for the imaginary part only, and not for the real part. We also confirm that the discrepancy cannot be removed by increasing the system size.

\section{Conclusions}

We find that the behaviours of the magnetization and free energy in the $\mathcal{PT}$-broken regime are markedly different from those in the case where $\mathcal{PT}$-symmetry is unbroken. Our results suggest that, when $\mathcal{PT}$ symmetry is spontaneously broken, the impurity is not screened in  the infrared. This lack of Kondo screening  is further suggested by a jump of the magnetisation associated with the $\mathcal{PT}$ phase transition at half-filling, as in the non-Hermitian Kondo model studied in \cite{ptkondo}. 

Meanwhile, our study of the Loschmidt echo shows that the overlap at large times decays as $t^{-1/2}$ whatever the regime. This decay is usually associated with healing - i.e., a situation where the two wires are fully connected \cite{St_phan_2011}.  In other words, we seem to find that, at low energy,  the impurity is not screened {\sl and} the wires are healed. We do not really understand how this can be made compatible with the expected circular behaviour of the RG explored in particular in  \cite{doi:10.1142/S0217751X93002265,ZAMOLODCHIKOV1994436,SALEUR1994205}. Of course, this circular behaviour is under control only in the perturbative regime (close to the isotropic Kondo problem), and it may be that it disappears in the RLM limit: we hope to get back to this question soon. In particular, we find the fact that equilibrium quantities in the $\mathcal{PT}$-broken regime can be calculated by analytical continuation from those in the unbroken regime very encouraging, since a lot is known about (UV) perturbative expansions of physical quantities in the (anisotropic) Kondo problem and boundary sine-Gordon problem. 

We also studied the Loschmidt echo starting from the initially disconnected chains. This requires further study: to our knowledge, it was not even known that this echo would obey universal behaviour in the different regimes, with different universal curves.

Finally, we note that  the quench protocol we examined in this paper was in fact already studied in the context of quantum impurity systems \cite{Latta2011,RevModPhys.85.79}. It therefore might also be possible to study a local quench in PTRLM experimentally by performing a quench to a $\mathcal{PT}$-symmetric optical trap.
\section{Acknowledgements}
TY thanks B. Doyon and R. Pereira, and O. Shpielberg for useful discussions. TY acknowledges the financial support from Takenaka Scholarship Foundation, and hospitality at the Institut de Physique Th\'eoreque, CEA, where the present work was initiated.
KB is grateful to G. Misguich, A. Michailidis and O. Gamayun for valuable suggestions. KB acknowledges the support from the European Research Council under the Starting Grant No. 805252 LoCoMacro. HS wishes to thank N. Andrei for discussions. This research was supported in part by the ERC Advanced Grant NuQFT.

\bibliography{pt-rlm.bib}

\appendix
\section{Non-hermitian quantum mechanics and biorthogonal basis}\label{biortho}
In general, when a given Hamiltonian $H$ is not hermitian, its eigenstates are not orthogonal. This in turn results in the loss of ordinary properties that hermitian quantum systems have, such as the orthogonality of wave functions and positivity of inner products. This problem can be in fact circumvented by properly extending the notion of basis to the biorthogonal basis $\{|R_n\rangle, |L_n\rangle\}\in \mathcal{H}\otimes\mathcal{H}^*$, where $\mathcal{H}$ and $\mathcal{H}^*$ are the Hilbert space and its dual space respectively, satisfying
\begin{equation}
    H|R_n\rangle=E_n|R_n\rangle,\quad H^\dagger|L_n\rangle=E^*_n|L_n\rangle.
\end{equation}
Biorthogonality further entails
\begin{equation}\label{biorth}
    \langle R_n|L_m\rangle=\delta_{n,m},\quad \sum_n|R_n\rangle\langle L_n|=\mathbb{I}.
\end{equation}
The trace of any observable over $\mathcal{H}$ can be expanded with respect to the biorthogonal basis. This can be best seen by first expand $\mathrm{Tr}\,\mathcal{O}$ with respect to a generic orthonormal basis $|\xi\rangle$ (which is always guaranteed to exist in a Hilbert space) as $\mathrm{Tr}\,\mathcal{O}=\sum_\xi\langle \xi|\mathcal{O}|\xi\rangle$. Inserting the spectral decomposition $\sum_n|R_n\rangle\langle L_n|=\mathbb{I}$, we then have
\begin{align}
    \mathrm{Tr}\,\mathcal{O}&=\sum_\xi\sum_n\langle L_n|\mathcal{O}|\xi\rangle\langle \xi|R_n\rangle \n
        &=\sum_n\langle L_n|\mathcal{O}|R_n\rangle.
\end{align}
Now suppose that there is a parity-operator $\mathcal{P}$ such that
\begin{equation}
    H^\dagger=\mathcal{P}H\mathcal{P},
\end{equation}
and
\begin{equation}
    \mathcal{P}|R_n\rangle=p_n|L_n\rangle
\end{equation}
with $\mathcal{P}^2=\mathbb{I}$ and $p_n=\pm1$. It is then customary to define an operator called $\mathcal{C}$-operator
\begin{equation}
    \mathcal{C}=\sum_np_n|R_n\rangle\langle L_n|
\end{equation}
so that $\rho=\mathcal{PC}=\sum_n|L_n\rangle\langle L_n|$ (not to be confused with density operator) which relates $H$ and $H^\dagger$ as
\begin{equation}
    H^\dagger\rho=\rho H.
\end{equation}
With this operator, we can provide an alternative way of interpreting the biorthogonal basis. Namely, the use of the biorthogonal basis is equivalent to working with the Hilbert space endowed with the metric $\rho$. For a given state $|\alpha\rangle=\sum_n\alpha_n|R_n\rangle\in\mathcal{H}$, let us define the {\it associate state} as
\begin{equation}\label{associated}
    \langle \Tilde{\alpha}|=\sum_n\alpha^*_n\langle L_n|\in\mathcal{H}^*.
\end{equation}
The inner product of two states $|\alpha\rangle$ and $|\beta\rangle$ for a biorthogonal system is then defined by
\begin{equation}\label{binorm1}
    \langle\beta|\alpha\rangle_\mathrm{PT}:=\langle\Tilde{\beta}|\alpha\rangle=\sum_n\beta^*_n\alpha_n.
\end{equation}
This inner product can be viewed as a conventional quantum mechanical inner product with the metric $\rho$ defined as follows
\begin{equation}
\langle\beta|\alpha\rangle_\rho:=\langle\beta|\rho|\alpha\rangle=\sum_n\beta^*_n\alpha_n,
\end{equation}
hence the equivalence of two point of views. It is readily seen that this inner product is positive definite 
Having discussed the states in a biorthogonal basis, we let us turn to observables. In terms of a given biorthogonal basis $\{|R_n\rangle, |L_n\rangle\}$, we can expand a generic operator $A$ as
\begin{equation}
    A=\sum_{n,m}a_{n m}|R_n\rangle\langle L_m|.
\end{equation}
The expectation value of $A$ in a pure state $|\psi\rangle=\sum_nc_n|R_n\rangle$, which we denote $\langle A\rangle_\psi$ is then defined by
\begin{equation}
    \langle A\rangle_\psi=\frac{\langle \Tilde{\psi}|A|\psi\rangle}{\langle \Tilde{\psi}|\psi\rangle}=\sum_{n,m}\frac{c^*_nc_ma_{nm}}{\sum_nc^*_nc_n},
\end{equation}
where we recall that  $\langle\Tilde{\psi}|=\sum_nc^*_n\langle L_n|$. Likewise, an arbitrary density operator $\varrho$ can be decomposed as
\begin{equation}
    \varrho=\sum_{n,m}\varrho_{n m}|R_n\rangle\langle L_m|.
\end{equation}
The statistical average of an observable $A$ with respect to $\varrho$ then reads
\begin{equation}
    \langle A\rangle=\mathrm{Tr}\,(\varrho A)=\sum_n\langle L_n|\varrho A|R_n\rangle=\sum_{n,m}a_{nm}\varrho_{mn}.
\end{equation}
Next, we turn to time-evolution. A pure state $|\psi\rangle$ is expected to time-evolve according to the following Schrodinger equation as in hermitian systems:
\begin{equation}
    \ii\frac{\partial}{\partial t}|\psi\rangle=H|\psi\rangle.
\end{equation}
Suppose that we start with an initial pure state $|\psi\rangle_0=\sum_n c_n|R_n\rangle$. The time-evolved state $|\psi(t)\rangle$ is given by
\begin{equation}
    |\psi(t)\rangle=\sum_n c_ne^{-\ii E_nt}|R_n\rangle
\end{equation}
 with the corresponding bra state
\begin{equation}\label{tevolleft}
    \langle \Tilde{\psi}(t)|=\sum_n c^*_ne^{\ii E^*_nt} \langle L_n|.
\end{equation}
The biorthogonal norm of the state $|\psi(t)\rangle$ then reads
\begin{equation}\label{normtdep}
    \langle \Tilde{\psi}(t)|\psi(t)\rangle=\sum_nc^*_nc_ne^{-\ii(E_n-E^*_n)t}.
\end{equation}
We observe that the norm is time-independent when the spectrum is entirely real, i.e. $E_n\in \mathbb{R}$ for all $n$. Furthermore, if the spectrum is real, then the associated state $|\Tilde{\psi}(t)\rangle=\sum_n c_ne^{-\ii E_nt} |L_n\rangle$, which is the conjugate of \eqref{tevolleft}, satisfies another Schrodinger equation $\ii\frac{\partial}{\partial t}|\Tilde{\psi}\rangle=H^\dagger|\Tilde{\psi}\rangle$, implying that the Heisenberg picture is still valid when $\mathcal{PT}$-symmetry is unbroken:
\begin{equation}
     \langle \Tilde{\psi}(t)|A|\psi(t)\rangle= \langle \Tilde{\psi}|e^{\ii Ht}Ae^{-\ii Ht}|\psi\rangle=\langle \Tilde{\psi}|A(t)|\psi\rangle
\end{equation}
with $A(t)=e^{\ii Ht}Ae^{-\ii Ht}$. We however note that the validity of the Heisenberg picture breaks down when at least one eigenvalue is complex, as the associated state $|\Tilde{\psi}(t)\rangle$ no longer follows the Schrodinger equation, i.e. $\ii\frac{\partial}{\partial t}|\Tilde{\psi}\rangle\neq H^\dagger|\Tilde{\psi}\rangle$ \cite{biorth}.

\begin{widetext}
\section{Details on the dot density}\label{dotdensity}
\subsection{Dot density at $T=0$}
 \subsubsection{Phase I}
 We first note that $k_\mathrm{F}=\arccos(-\frac{\mu}{2})\geq\frac{\pi}{2}$. Then we can safely decompose the integral as
\begin{align}\label{anacont1}
    d=\int_0^{k_\mathrm{F}}\frac{\dd k}{\pi}\frac{\tgamma\sin^2k}{\tgamma^2+(1-2\tgamma)\cos^2k}&=\int_0^{\pi/2}\frac{\dd k}{\pi}\frac{\tgamma\sin^2k}{\tgamma^2+(1-2\tgamma)\cos^2k}+\lim_{\varepsilon\to0^+}\int_{\pi/2+\varepsilon}^{k_\mathrm{F}}\frac{\dd k}{\pi}\frac{\tgamma\sin^2k}{\tgamma^2+(1-2\tgamma)\cos^2k}\n
    &=\frac{1}{2}+\frac{\tgamma}{1-2\tgamma}\int_{-\infty}^{\tan k_\mathrm{F}}\frac{\dd x}{\pi}\Bigg(\frac{1}{1+\frac{\tgamma^2}{(1-\tgamma)^2}x^2}-\frac{1}{1+x^2}\Bigg)\n
    &=\frac{1}{2}+\frac{\tgamma}{1-2\tgamma}\bigg(\int_0^{\infty}\frac{\dd x}{\pi}-\int_0^{-\tan k_\mathrm{F}}\frac{\dd x}{\pi}\bigg)\Bigg(\frac{1}{1+\frac{\tgamma^2}{(1-\tgamma)^2}x^2}-\frac{1}{1+x^2}\Bigg)\n
    &=\frac{1}{2}+\frac{1-\tgamma}{1-2\tgamma}\int_{u(\tgamma)}^{\infty}\frac{\dd x}{\pi}\frac{1}{1+x^2}-\frac{\tgamma}{1-2\tgamma}\int_{-\tan k_\mathrm{F}}^{\infty}\frac{\dd x}{\pi}\frac{1}{1+x^2}\n
    &=1+\frac{1-\tgamma}{\pi u(\tgamma)(1-2\tgamma)}\sum_{n=1}^\infty\frac{(-1)^n}{2n-1}u(\tgamma)^{2n}+\frac{\tgamma}{\pi \tan k_\mathrm{F}(1-2\tgamma)}\sum_{n=1}^\infty\frac{(-1)^n}{2n-1}\tan^{2n}k_\mathrm{F}
\end{align}
where we remarked $\tan k_\mathrm{F}<0$, and defined $u(\tgamma)=-\tgamma\tan k_\mathrm{F}/(1-\tgamma)=\tgamma\sqrt{4-\mu^2}/(\mu(1-\tgamma))$. We also assumed that $u(\tgamma)<1$ for the time being. The point here is that, we cannot naively change the integration variable $x=\tan k$ in the second equality as the integration region $[0,k_\mathrm{F}]$ does not one-to-one correspond to $[0,\tan k_\mathrm{F}]$. 
We then have the UV expansion 
\begin{equation}
    d=1-\frac{1}{\pi(1-2\tgamma)}\big[(1-\tgamma)\arctan u(\tgamma)-\tgamma\arctan \frac{\sqrt{4-\mu^2}}{\mu}\big].
\end{equation}
where the function $\arctan$ is defined by its series expansion (this point has to be made clear since multiple definitions of the function $\arctan z={1\over 2i}\ln {1+iz\over 1-iz}$ play a role in the discussion of analytical continuations in the body of this paper):
\begin{equation}
\arctan x=\sum_{0}^\infty (-1)^{2k+1} {x^{2k+1}\over 2k+1},~~|x|<1
\end{equation}
For the opposite case $u(\tgamma)>1$, one can perform the dual transformation $x\mapsto1/x$ in \eqref{anacont1} and get the IR expansion
\begin{align}
    d&=\frac{1}{2}+\frac{1-\tgamma}{1-2\tgamma}\int_0^{1/u(\tgamma)}\frac{\dd x}{\pi}\frac{1}{1+x^2}-\frac{\tgamma}{1-2\tgamma}\int_0^{-1/\tan k_\mathrm{F}}\frac{\dd x}{\pi}\frac{1}{1+x^2} \n
    &=\frac{1}{2}-\frac{u(\tgamma)(1-\tgamma)}{\pi (1-2\tgamma)}\sum_{n=1}^\infty\frac{(-1)^n}{2n-1}u(\tgamma)^{-2n}-\frac{\tan k_\mathrm{F}\tgamma}{\pi(1-2\tgamma)}\sum_{n=1}^\infty\frac{(-1)^n}{2n-1}\left(\frac{1}{\tan k_\mathrm{F}}\right)^{2n}\n
    &=\frac{1}{2}+\frac{1}{\pi(1-2\tgamma)}\big[(1-\tgamma)\arctan\frac{1}{u(\tgamma)}-\tgamma\arctan\frac{\mu}{\sqrt{4-\mu^2}}\big].
\end{align}
Therefore, in summary,
\begin{equation}
    d=\begin{dcases}
    1-\frac{1}{\pi(1-2\tgamma)}\big[(1-\tgamma)\arctan u(\tgamma)-\tgamma\arctan \frac{\sqrt{4-\mu^2}}{\mu}\big] & u(\tgamma)<1\\
    \frac{1}{2}+\frac{1}{\pi(1-2\tgamma)}\big[(1-\tgamma)\arctan\frac{1}{u(\tgamma)}-\gamma^2_1\arctan\frac{\mu}{\sqrt{4-\mu^2}}\big]& u(\tgamma)>1
    \end{dcases},
\end{equation}
and in particular, under the scaling limit $\tgamma,\mu\ll1$, we have $u(\tgamma)\simeq \tgamma/\mu$, and 
\begin{equation}
    d=\begin{dcases}
    1-\frac{1}{\pi}\arctan u(\tgamma) & u(\tgamma)<1\\
    \frac{1}{2} +\frac{1}{\pi}\arctan\big(\frac{1}{u(\tgamma)}\big)& u(\tgamma)>1
    \end{dcases}.
\end{equation}
\subsubsection{Phase II}
We can calculate the dot density in the phase II in the same way for $\mu>0$.
\begin{align}
    d&=\int_0^{k_\mathrm{F}}\frac{\dd k}{\pi}\frac{\tgamma\sin^2k}{\tgamma^2+(1-2\tgamma)\cos^2k}+\frac{1-\tgamma}{1-2\tgamma}\n
    &=\int_0^{\pi/2}\frac{\dd k}{\pi}\frac{\tgamma\sin^2k}{\tgamma^2+(1-2\tgamma)\cos^2k}+\frac{1-\tgamma}{1-2\tgamma}+\lim_{\varepsilon\to0^+}\int_{\pi/2+\varepsilon}^{k_\mathrm{F}}\frac{\dd k}{\pi}\frac{\tgamma\sin^2k}{\tgamma^2+(1-2\tgamma)\cos^2k}\n
    &=\frac{1}{2}+\lim_{\varepsilon\to0^+}\int_{\pi/2+\varepsilon}^{k_\mathrm{F}}\frac{\dd k}{\pi}\frac{\tgamma\sin^2k}{\tgamma^2+(1-2\tgamma)\cos^2k}\n
    &=\frac{1}{2}-\frac{1-\tgamma}{1-2\tgamma}\int_{v(\tgamma)}^{\infty}\frac{\dd x}{\pi}\frac{1}{1+x^2}-\frac{\tgamma}{1-2\tgamma}\int_{-\tan k_\mathrm{F}}^{\infty}\frac{\dd x}{\pi}\frac{1}{1+x^2}\n
    &=\frac{\tgamma}{2\tgamma-1}-\frac{1-\tgamma}{\pi v(\tgamma)(1-2\tgamma)}\sum_{n=1}^\infty\frac{(-1)^n}{2n-1}v(\tgamma)^{2n}-\frac{\tgamma}{\pi \tan k_\mathrm{F}(1-2\tgamma)}\sum_{n=1}^\infty\frac{(-1)^n}{2n-1}\tan^{2n}k_\mathrm{F},
\end{align}
where we assumed $v(\tgamma):=-u(\tgamma)<1$. Note that when passing from the second line to the third line, we used
\begin{equation}\label{dotidentity}
   \int_0^{\pi/2}\frac{\dd k}{\pi}\frac{\tgamma\sin^2k}{\tgamma+(1-2\tgamma)\cos^2k}+\frac{1-\tgamma}{1-2\tgamma}=\frac{1}{2}.
\end{equation}
The UV expansion in the phase II then reads
\begin{equation}
    d=\frac{\tgamma}{2\tgamma-1}+\frac{1}{\pi(2\tgamma-1)}\big[(\tgamma-1)\arctan v(\tgamma)-\tgamma\arctan \frac{\sqrt{4-\mu^2}}{\mu}\big].
\end{equation}
As in the phase I, the dual transformation in the integrations instead gives the IR expansion
\begin{align}
    d&=\frac{1}{2}+\frac{v(\tgamma)(\tgamma-1)}{\pi (2\tgamma-1)}\sum_{n=1}^\infty\frac{(-1)^n}{2n-1}v(\tgamma)^{-2n}+\frac{\tgamma}{\pi(2\tgamma-1)}\arctan\frac{\mu}{\sqrt{4-\mu^2}}\n
    &=\frac{1}{2}-\frac{1}{\pi(2\tgamma-1)}\big[(\tgamma-1)\arctan\frac{1}{v(\tgamma)}-\tgamma\arctan\frac{\mu}{\sqrt{4-\mu^2}}\big].
\end{align}
\subsubsection{Phase III}
A similar analysis can be carried out in the phase III and for $\mu>0$
\begin{align}
   d&=\int_0^{k_\mathrm{F}}\frac{\dd k}{\pi}\frac{\tgamma\sin^2k}{\tgamma^2+(1-2\tgamma)\cos^2k}+\frac{2-2\tgamma}{1-2\tgamma}\n
  &=\int_0^{\pi/2}\frac{\dd k}{\pi}\frac{\tgamma\sin^2k}{\tgamma+(1-2\tgamma)\cos^2k}+\frac{2-2\tgamma}{1-2\tgamma}+\lim_{\varepsilon\to0^+}\int_{\pi/2+\varepsilon}^{k_\mathrm{F}}\frac{\dd k}{\pi}\frac{\tgamma\sin^2k}{\tgamma^2+(1-2\tgamma)\cos^2k}\n
  &=\frac{1}{2}+\frac{1-\tgamma}{1-2\tgamma}+\lim_{\varepsilon\to0^+}\int_{\pi/2+\varepsilon}^{k_\mathrm{F}}\frac{\dd k}{\pi}\frac{\tgamma\sin^2k}{\tgamma^2+(1-2\tgamma)\cos^2k}\n
  &=\frac{1}{2}+\frac{1-\tgamma}{1-2\tgamma}-\frac{1-\tgamma}{1-2\tgamma}\int_{-u(\tgamma)}^{\infty}\frac{\dd x}{\pi}\frac{1}{1+x^2}-\frac{\tgamma}{1-2\tgamma}\int_{-\tan k_\mathrm{F}}^{\infty}\frac{\dd x}{\pi}\frac{1}{1+x^2}\n
    &=1-\frac{1-\tgamma}{\pi v(\tgamma)(1-2\tgamma)}\sum_{n=1}^\infty\frac{(-1)^n}{2n-1}v(\tgamma)^{2n}-\frac{\tgamma}{\pi \tan k_\mathrm{F}(1-2\tgamma)}\sum_{n=1}^\infty\frac{(-1)^n}{2n-1}\tan^{2n}k_\mathrm{F},
\end{align}
where we again used \eqref{dotidentity} in the third line, which is still valid in the phase III.
We therefore have the UV expansion in the phase III
\begin{equation}
    d=1+\frac{1}{\pi(1-2\tgamma)}\big[(1-\tgamma)\arctan v(\tgamma)+\tgamma\arctan \frac{\sqrt{4-\mu^2}}{\mu}\big].
\end{equation}
Likewise, the IR expansion ($v(\tgamma)>1$) reads
\begin{align}
    d&=\frac{1}{2}+\frac{1-\tgamma}{1-2\tgamma}-\frac{1-\tgamma}{1-2\tgamma}\int_0^{1/v(\tgamma)}\frac{\dd x}{\pi}\frac{1}{1+x^2}-\frac{\tgamma}{1-2\tgamma}\int_0^{-1/\tan k_\mathrm{F}}\frac{\dd x}{\pi}\frac{1}{1+x^2}\n
    &=\frac{1}{2}+\frac{1-\tgamma}{1-2\tgamma}-\frac{1}{\pi(1-2\tgamma)}\big[(1-\tgamma)\arctan \frac{1}{v(\tgamma)}+\tgamma\arctan\frac{\mu}{\sqrt{4-\mu^2}}\big].
\end{align}

\subsection{Proof of $d=\frac{1}{2}$ at half-filling}
Defining $D(k)=\Tilde{\Gamma}\sin^2k/(\Tilde{\Gamma}^2+(1-2\Tilde{\Gamma})\cos^2k)$, we rewrite \eqref{dot2} as
\begin{equation}
    d=\int_0^{\pi/2} \frac{\dd k}{\pi}\frac{1}{1+e^{-2\beta \cos k}}D(k)+\int_{\pi/2}^\pi\frac{\dd k}{\pi}\frac{e^{2\beta\cos k}}{1+e^{2\beta \cos k}}D(k).
\end{equation}
In each integration region, the Fermi distribution can be expanded as
\begin{align}
    \frac{1}{1+e^{-2\beta \cos k}}&=\sum_{n=0}^\infty
\Big(-e^{-2\beta\cos k}\Big)^n=\sum_{n=0}^\infty(-1)^n\sum_{m=0}^\infty\frac{(-2n\beta)^m}{m!}\cos^mk\n
\frac{e^{2\beta \cos k}}{1+e^{2\beta \cos k}}&=\sum_{n=0}^\infty
(-1)^ne^{2\beta(n+1)\cos k}=\sum_{n=0}^\infty(-1)^n\sum_{m=0}^\infty\frac{[2(n+1)\beta]^m}{m!}\cos^mk.
\end{align}
As we will see later, it is useful to divide the case $0<\Tilde{\Gamma}\leq 1/2$ and $1/2\leq \Tilde{\Gamma}<1$ and analyze separately. Let us deal with the former case first in which case we can expand $D(k)$ as
\begin{equation}
    D(k)=\frac{1}{(\Tilde{\Gamma}-1)^2+(2\Tilde{\Gamma}-1)\sin^2k}=\frac{1}{(\Tilde{\Gamma}-1)^2}\sum_{l=0}^\infty\Big(-\frac{2\Tilde{\Gamma}-1}{(\Tilde{\Gamma}-1)^2}\Big)^l\sin^{2l}k,
\end{equation}
where we noted that $-1< -(2\Tilde{\Gamma}-1)/(\Tilde{\Gamma}-1)^2\leq0$.
We then perform the integrations each of which reads
\begin{align}
     \int_0^{\pi/2}\dd k\cos^mk\sin^{2(l+1)}k&=\frac{1}{2}B\Big(\frac{m+1}{2},\frac{2l+3}{2}\Big)=\frac{1}{2}\frac{\Gamma\big(\frac{m+1}{2}\big)\Gamma\big(l+\frac{3}{2}\big)}{\Gamma\big(l+\frac{m}{2}+2\big)}\n
      \int_{\pi/2}^\pi\dd k\cos^mk\sin^{2(l+1)}k&=\frac{(-1)^m}{2}\frac{\Gamma\big(\frac{m+1}{2}\big)\Gamma\big(l+\frac{3}{2}\big)}{\Gamma\big(l+\frac{m}{2}+2\big)}.
\end{align}
Using them, we can express the dot density $d$ as
\begin{align}
    d&=\frac{\Tilde{\Gamma}}{2\pi(\Tilde{\Gamma}-1)^2}\sum_{n=0}^\infty(-1)^n\sum_{m=0}^\infty\frac{(-2n\beta)^m+(-2(n+1)\beta)^m}{m!}\sum_{l=0}^\infty\Big(-\frac{2\Tilde{\Gamma}-1}{(\Tilde{\Gamma}-1)^2}\Big)^l\frac{\Gamma\big(\frac{m+1}{2}\big)\Gamma\big(l+\frac{3}{2}\big)}{\Gamma\big(l+\frac{m}{2}+2\big)}\n
    &=\frac{\Tilde{\Gamma}}{2\pi(\Tilde{\Gamma}-1)^2}\sum_{l=0}^\infty\Big(-\frac{2\Tilde{\Gamma}-1}{(\Tilde{\Gamma}-1)^2}\Big)^l\frac{\Gamma\big(\frac{1}{2}\big)\Gamma\big(l+\frac{3}{2}\big)}{\Gamma\big(l+2\big)}\n
    &=\frac{\Tilde{\Gamma}}{4(\Tilde{\Gamma}-1)^2}\,{}_2F_1\Big(1,\frac{3}{2};2;-\frac{2\Tilde{\Gamma}-1}{(\Tilde{\Gamma}-1)^2}\Big). \label{hfdot2}
\end{align}
Since ${}_2F_1(1,\frac{3}{2};2;z)=(\frac{1}{2}+\frac{1}{2}\sqrt{1-z})^{-1}/\sqrt{1-z}$ (note the {\it positive} branch of square root), it follows that $d=\frac{1}{2}$. For the another case $1/2\leq \Tilde{\Gamma}<1$, noticing that $-1<(1-2\Tilde{\Gamma})/\Tilde{\Gamma}^2\leq0$, we can use another expansion
\begin{equation}
    D(k)=\frac{1}{\Tilde{\Gamma}^2}\sum_{l=0}^\infty\Big(\frac{1-2\Tilde{\Gamma}}{\Tilde{\Gamma}^2}\Big)^l\cos^{2l}k,
\end{equation}
yielding the same result.

\section{Loschmidt echo}\label{echosec}
Here we elucidate the computations of the Loschmidt echo, which is defined by
\begin{equation}\label{echo}
    \mathcal{L}(t)={}_0\langle \mathrm{GS}|e^{\ii tH_0}e^{-\ii tH}|\mathrm{GS}\rangle_0=e^{\ii tE_\mathrm{GS}}{}_0\langle \mathrm{GS}|e^{-\ii tH}|\mathrm{GS}\rangle_0,
\end{equation}
where $|GS\rangle_0$ denotes the ground state of a two disconnected chains. We consider the case where an arbitrary chemical potential is imposed. Let us first expand $|\mathrm{GS}\rangle_0$ in terms of the mode operator in the PTRLM.  Defining the cut-off mode $M=\lfloor \frac{N+2}{2\pi}k_\mathrm{F}\rfloor$ with the Fermi momentum $k_\mathrm{F}=\arccos \left(-\frac{\mu}{2}\right)$,
 \begin{align}
     |\mathrm{GS}\rangle_0=\prod_{i\in\mathrm{FS}}c^\dagger_{k_i}|0\rangle=\prod_{i\in\mathrm{FS}}\sum_{x_i}\phi(k_i;x_i)c^\dagger_{x_i}|0\rangle&=\prod_{i\in\mathrm{FS}}\sum_{x_i}\sum_{l_{i}}\phi(k_i;x_i)L^*(l_{i};x_i)c^\dagger_{R,l_{i}}|0\rangle\n
     &=\prod_{i\in\mathrm{FS}}\sum_{l_{i}}\mathtt{L}^*(k_i,l_i)c^\dagger_{R,l_{i}}|0\rangle,
 \end{align}
 where we defined $\mathtt{L}^*(k_i,l_i):=\sum_{x}\phi(k_i;x)L^*(l_{i};x)$. In the same way, we also expand ${}_0\langle \mathrm{GS}|$ as
 \begin{equation}
     {}_0\langle \mathrm{GS}|=\langle 0|\prod_{i\in\mathrm{FS}}\sum_{l_{i}}\mathtt{R}(k_i,l_i)c_{L,l_{i}}.
 \end{equation}
 Now, inserting the biorthogonal resolution of identity $1=\sum_{K}\frac{1}{K!}\prod_{j=1}^K\sum_{m_j}c^\dagger_{R,m_1}\cdots c^\dagger_{R,m_K}|0\rangle\langle 0|c_{L,m_K}\cdots c_{L,m_1}$ into \eqref{echo}, we obtain
 \begin{align}
     \mathcal{L}(t)&=\frac{e^{\ii tE_\mathrm{GS}}}{(2M+1)!}\prod_{i\in\mathrm{FS}}\sum_{m_i}e^{-\ii E_{m_i}t}\det_{i,j\in\mathrm{FS}}\mathtt{L}^*(k_i,m_j)\det_{i',j'\in\mathrm{FS}}\mathtt{R}(k_{i'},m_{j'})\n
     &=e^{\ii tE_\mathrm{GS}}\det_{i,j\in\mathrm{FS}}\mathtt{M}(k_i;k_j),
 \end{align}
 where $\mathtt{M}(k_i,k_j;t)=\sum_{m}e^{-\ii E_{m}t}\mathtt{L}^*(k_i,m)\mathtt{R}(k_{j},m)$. To proceed, we divide $\mathtt{M}(k_i,k_j;t)$ into $\mathtt{M}(k_i,k_j)=\mathtt{M}_+(k_i,k_j)+\mathtt{M}_-(k_i,k_j)$ each of which is defined as $\mathtt{M}_\pm(k_i,k_j;t)=\sum_{m_\pm}e^{-\ii E_{m_\pm}t}\mathtt{L}^*(k_i,m_\pm)\mathtt{R}(k_{j},m_\pm)$.
 \subsection{Phase I}
 Let us start with $\mathtt{M}_+(k_i,k_j;t)$. Note that $\mathtt{L}^*(k,m_+)$ and $\mathtt{R}(k,m_+)$ can be explicitly evaluated as
 \begin{align}
    \mathtt{L}^*(k;m_+)=\sum_x\phi(k;x)L^*_+(m_+;x)&=2\mathcal{N}^*_{\mathrm{L},+}(m_+)\sqrt{\frac{1}{N+2}}\sum_{x>0}(\vartheta(k)+\mathcal{M}_+\vartheta(-k))\sin kx\cos[m_+(x-\delta_+(m_+))]\n
    &=\frac{\mathcal{N}^*_{\mathrm{L},+}(m_+)}{\sqrt{N+2}}(\vartheta(k)+\mathcal{M}_+\vartheta(-k))\frac{\sin k\cos[m_+\delta_+(m_+)]}{\cos m_+-\cos k},
 \end{align}
 and
 \begin{equation}
     \mathtt{R}(k;m_+)=\frac{\mathcal{N}_{\mathrm{R},+}(m_+)}{\sqrt{N+2}}(\vartheta(k)+\mathcal{M}_+\vartheta(-k))\frac{\sin k\cos[m_+\delta_+(m_+)]}{\cos m_+-\cos k}.
 \end{equation}
 Recalling that $\cos^2[m_+\delta_+(m_+)]=4\Tilde{\Gamma}^2\sin^2m/\Delta_+(m)$, $\mathtt{M}_+(k_i,k_j;t)$ can therefore be expressed as
 \begin{equation}\label{buildm}
 \mathtt{M}_+(k_i,k_j;t)=\frac{4\Tilde{\Gamma}^2\sin k_i\sin k_j}{N+2}(\vartheta(k_i)+\mathcal{M}_+\vartheta(-k_i))(\vartheta(k_j)+\mathcal{M}_+\vartheta(-k_j))\sum_{m_+}\frac{\hat{\mathcal{N}}_+(m_+)\sin^2m_+e^{-\ii t E_{m_+}}}{\Delta_+(m_+)(\cos m_+-\cos k_i)(\cos m_+-\cos k_j)}.
 \end{equation}
 We focus on computing the building block
 \begin{equation}\label{qplus}
    \mathtt{Q}_+(k_i,k_j;t) :=\sum_{m_+>0}w_1(m_+;k_i,k_j;t), \quad w_1(m_+;k_i,k_j;t)= \frac{\hat{\mathcal{N}}_+(m_+)\sin^2m_+e^{-\ii t E_{m_+}}}{\Delta_+(m_+)(\cos m_+-\cos k_i)(\cos m_+-\cos k_j)}.
 \end{equation}
In the thermodynamic limit, the calculation of \eqref{qplus} is apparently plagued by poles, thereby necessitates a systematic way to deal with them. To this end, we introduce a counting function $Q_\sigma(k)$ that satisfies
\begin{equation}
    Q_\sigma(k)=kN+\ii\log S_\sigma(k),
\end{equation}
where the S-matrix is $S_\sigma(k)=(\Gamma_\sigma-e^{-2\ii k})/(\Gamma_\sigma-e^{2\ii k})$. Only $k$ such that $Q_\sigma(k)=2\pi n$ for some quantum number $n\in\mathbb{Z}$ is allowed by the quantization condition \eqref{quantcond}. Using $Q(k)$, we can rewrite the sum over $k$ (suppose that $k$ satisfies the quantization condition with respect to the sign $\sigma$) as
\begin{equation}
    \sum_{k}\longrightarrow\sum_n\oint_{\mathcal{C}_n}\frac{\dd k}{2\pi}\frac{- Q_\sigma'(k)}{e^{-\ii Q_\sigma(k)}-1},
\end{equation}
where $\mathcal{C}_n$ is a closed contour encircling each pole. We then rewrite $ \mathtt{Q}_+(k_i,k_j;t)$ as
 \begin{align}
      \mathtt{Q}_+(k_i,k_j;t)&=\left(\int_{0-\ii\epsilon}^{\pi-\ii\epsilon}-\int_{0+\ii\epsilon}^{\pi+\ii\epsilon}\right)\frac{\dd m}{2\pi}\frac{-Q'_+(m)}{e^{-\ii Q_+(m)}-1}w_1(m;k_i,k_j;t) \n
     &\quad -\ii\sum_{l}\left.\mathrm{Res}\right|_{m_+=\Tilde{k}_l}\frac{-Q'_+(m_+)}{e^{-\ii Q_+(m_+)}-1}w_1(m_+;k_i,k_j;t),
 \end{align}
where $\Tilde{k}_j>0$ are the possible poles in $w_1(m_+;k_i,k_j;t)$. Noticing that $\hat{\mathcal{N}}_+(m)Q'_+(m)=2(\gamma^*)^2/\Tilde{\Gamma}$, we can further recast it into the following form
 \begin{equation}
     \mathtt{Q}_+(k_i,k_j;t)=\frac{2\ii(\gamma^*)^2}{\Tilde{\Gamma}}(\mathtt{I}_{ij}(t)+\mathtt{W}(k_i,k_j;t)),
 \end{equation}
 where
 \begin{equation}
 \mathtt{I}_{ij}(t)=\ii\left(\int_{0-\ii\epsilon}^{\pi-\ii\epsilon}-\int_{0+\ii\epsilon}^{\pi+\ii\epsilon}\right)\frac{\dd m}{2\pi}w_2(m;k_i,k_j;t),\quad 
 \mathtt{W}(k_i,k_j;t)=\sum_l\left.\mathrm{Res}\right|_{m_+=\Tilde{k}_l}w_2(m_+;k_i,k_j;t)
 \end{equation}
 with
 \begin{equation}
     w_2(m_+;k_i,k_j;t)=\frac{\sin^2m_+e^{-\ii t E_{m_+}}}{\Delta_+(m_+)(e^{-\ii Q_+(m_+)}-1)(\cos m_+-\cos k_i)(\cos m_+-\cos k_j)}.
 \end{equation}
 Let us first analyze how $\mathtt{I}_{ij}(t)$ behaves for large $N$. Notice that the second term in $\mathtt{I}_{ij}(t)$ goes to zero when $N\to\infty$. Therefore only the first term remains finite in the thermodynamic limit. Its behavior distinctively differs depending on the value of $\tgamma$, and goes as
 \begin{align}
   \lim_{\epsilon\to0}\lim_{N\to\infty} \mathtt{I}_{ij}(t)&=\ii\left(\mathtt{J}(k_i,k_j;t)+\mathtt{J}\left(\frac{\pi}{2}-k_i,\frac{\pi}{2}-k_j;t\right)^*\right) \n
  &\quad +\vartheta\left(\frac{1}{2}-\tgamma\right)\frac{\ii(1-\tgamma)}{8|\tgamma|}\frac{e^{-tE_\mathrm{b}}}{(|\tgamma|+\ii\sqrt{1-2\tgamma}\cos k_i)(|\tgamma|+\ii\sqrt{1-2\tgamma}\cos k_j)},
   \end{align}
 where
 \begin{equation}
 \mathtt{J}(k_i,k_j;t)=\ii e^{2\ii t}\int_0^\infty\frac{\dd x}{2\pi}\frac{e^{-tx}\sqrt{-x^2+4\ii x}}{\left(4\tgamma^2+(1-2\tgamma)(2+\ii x)^2\right)(2+\ii x-2\cos k_i)(2+\ii x-2\cos k_j)}.
 \end{equation}
 Numerically it can be easily seen that the second term is dominant when $\tgamma<1/2$, and
 the first term, which is decaying algebraically, characterizes small ripples around the exponential decay. However the second term vanishes when $\tgamma<1/2$, and the dominant contribution is taken over by the first term, which for large $t$ behaves as
  \begin{equation}
   \ii\left(\mathtt{J}(k_i,k_j;t)+\mathtt{J}\left(\frac{\pi}{2}-k_i,\frac{\pi}{2}-k_j;t\right)^*\right)
    =\frac{\ii}{128\sqrt{\pi}(1-\tgamma)^2}\left(\frac{e^{\ii(2t+\pi/4)}}{\sin^2k_i\sin^2k_j}+\frac{e^{-\ii(2t+\pi/4)}}{\cos^2k_i\cos^2k_j}\right)t^{-\frac{3}{2}}+\mathcal{O}(t^{-2}).
  \end{equation}
 \noindent$\bullet$ $k_i\neq k_j>0$: (the usual trick) $ \mathtt{W}(k_i,k_j;t)$ becomes
 \begin{align}
     \mathtt{W}(k_i,k_j;t)&= (\left.\mathrm{Res}\right|_{m_+=k_i}+(\left.\mathrm{Res}\right|_{m_+=k_j})w_2(m_+;k_i,k_j;t)\n
     &=\mathtt{X}_{ij}(t)+\mathtt{X}_{ji}(t) \label{q1}
 \end{align}
 where
 \begin{equation}
     \mathtt{X}_{ij}(t)=\frac{1}{2}\left(\sin k_i-\ii\frac{(1-\Tilde{\Gamma})\cos k_i}{\Tilde{\Gamma}}\right)\frac{e^{-\ii E_{k_i}t}}{\Delta_+(k_i)(\cos k_i-\cos k_j)}
 \end{equation}
$\bullet$ $k_j<0<k_i,k_i\neq-k_j$: In this case 
$\mathtt{W}(k_i,k_j;t)$ can be calculated as
\begin{align}
     \mathtt{W}(k_i,k_j;t)&= (\left.\mathrm{Res}\right|_{m_+=k_i}+(\left.\mathrm{Res}\right|_{m_+=-k_j})w_2(m_+;k_i,k_j;t)\n
     &=\mathtt{X}_{ij}(t)+ \mathtt{Y}_{ji}(t).
 \end{align},
 where
 \begin{equation}
    \mathtt{Y}_{ji}(t)= -\frac{1}{2}\left(\sin k_j+\ii\frac{(1-\Tilde{\Gamma})\cos k_j}{\Tilde{\Gamma}}\right)\frac{e^{-\ii E_{k_j}t}}{\Delta_+(k_j)(\cos k_j-\cos k_i)}
 \end{equation}
$\bullet$ $k_i<0<k_j,k_i\neq-k_j$: Following the  same manipulation as above, $\mathtt{W}(k_i,k_j;t)$
\begin{align}
     \mathtt{W}(k_i,k_j;t)&= (\left.\mathrm{Res}\right|_{m_+=-k_i}+(\left.\mathrm{Res}\right|_{m_+=k_j})w_2(m_+;k_i,k_j;t)\n
     &=\mathtt{Y}_{ij}(t)+ \mathtt{X}_{ji}(t).
 \end{align}
$\bullet$ $k_i\neq k_j<0$: Likewise,
\begin{equation}
  \mathtt{W}(k_i,k_j;t)=\mathtt{Y}_{ij}(t)+\mathtt{Y}_{ji}(t).
\end{equation}
The remaining case is $|k_i|=|k_j|$, for which we have a double pole that can be treated by the same trick above.

$\bullet$ $k_i= k_j>0$: We can evaluate $ \mathtt{W}(k_i,k_i;t)$ as
\begin{equation}
     \mathtt{W}(k_i,k_i;t)=\left.\mathrm{Res}\right|_{m_+=k_i}w_2(m_+;k_i,k_i;t)= -\frac{\ii e^{-\ii E_{k_j}t}}{16\Tilde{\Gamma}^2\sin^2k_i}(N+2)+\mathtt{D}_i(t),
 \end{equation}
 where
 \begin{equation}
      \mathtt{D}_i(t)=\frac{e^{-\ii E_{k_j}t}}{\Delta_+(k_i)}\left(\sin k_i-\ii\frac{(1-\Tilde{\Gamma})\cos k_i}{\Tilde{\Gamma}}\right)\left(\ii t-\frac{4(1-2\Tilde{\Gamma})\cos k_i}{\Delta_+(k_i)}\right)-\frac{e^{-\ii E_{k_j}t}}{2\Delta_+(k_i)}\left(\frac{\cos k_i}{\sin k_i}+\ii \frac{1-\Tilde{\Gamma}}{\Tilde{\Gamma}}\right)
 \end{equation}
 $\bullet$ $0<k_i= -k_j$: In this case $ \mathtt{W}(k_i,-k_i;t)$ is the same as in the previous case, so $\mathtt{W}(k_i,-k_i;t)=\mathtt{D}_i(t)$.
 
\noindent$\bullet$ $0<k_j= -k_i$: Likewise
\begin{equation}
    \mathtt{W}(k_i,-k_i;t)=\left.\mathrm{Res}\right|_{m_+=-k_i}w_2(m_+;k_i,-k_i;t)= -\frac{\ii e^{-\ii E_{k_j}t}}{16\Tilde{\Gamma}^2\sin^2k_i}(N+2)+\mathtt{E}_i(t).
\end{equation}
where
\begin{equation}
   \mathtt{E}_i(t)=\frac{e^{-\ii E_{k_j}t}}{\Delta_+(k_i)}\left(\sin k_i+\ii\frac{(1-\Tilde{\Gamma})\cos k_i}{\Tilde{\Gamma}}\right)\left(-\ii t+\frac{4(1-2\Tilde{\Gamma})\cos k_i}{\Delta_+(k_i)}\right)+\frac{e^{-\ii E_{k_j}t}}{2\Delta_+(k_i)}\left(\frac{\cos k_i}{\sin k_i}-\ii \frac{1-\Tilde{\Gamma}}{\Tilde{\Gamma}}\right)
\end{equation}
$\bullet$ $k_i=k_j<0$: In this case $ \mathtt{W}(k_i,k_i;t)$ is the same as in the previous case, therefore $ \mathtt{W}(k_i,k_i;t)=\mathtt{E}_i(t)$.

To summary, we have
\begin{equation}
    \mathtt{W}(k_i,k_j;t)=\begin{dcases*}
    \mathtt{X}_{ij}(t)+\mathtt{X}_{ji}(t) & $k_i\neq k_j>0$\\
\mathtt{X}_{ij}(t)+\mathtt{Y}_{ji}(t) & $k_i>0>k_j$, $k_i\neq-k_j$\\
    \mathtt{Y}_{ij}(t)+\mathtt{X}_{ji}(t) & $k_j>0>k_i$, $k_i\neq-k_j$\\
    \mathtt{Y}_{ij}(t)+\mathtt{Y}_{ji}(t) & $k_i\neq k_j<0$\\
   -\frac{\ii e^{-\ii E_{k_j}t}}{16\Tilde{\Gamma}^2\sin^2k_i}(N+2)+\mathtt{D}_i(t) & $k_i=\pm k_j$, $k_i>0$ \\
  -\frac{\ii e^{-\ii E_{k_j}t}}{16\Tilde{\Gamma}^2\sin^2k_i}(N+2)+\mathtt{E}_i(t) &  $k_i=\pm k_j$, $k_i<0$ 
  \end{dcases*}.
\end{equation}
Therefore
\begin{equation}
    \mathtt{M}_+(k_i,k_j;t)=\frac{(\gamma^*)^2}{2\Tilde{\Gamma}}\begin{dcases*}
  1  & $k_i= k_j>0$ \\
  -\mathcal{M}_+ &  $k_i=-k_j$\\
  \mathcal{M}_+^2 &  $k_i=k_j<0$ \\
  0 & otherwise
  \end{dcases*}+\frac{2\pi}{N+2}\mathtt{A}_{ij}(t),
\end{equation}
where
\begin{equation}\label{amatrix}
   \mathtt{A}_{ij}(t)=\frac{8\ii(\gamma^*)^2\Tilde{\Gamma}\sin k_i\sin k_j}{2\pi}\begin{dcases*}
   \mathtt{I}_{ij}(t)+ \mathtt{X}_{ij}(t)+\mathtt{X}_{ji}(t) & $k_i\neq k_j>0$\\
\mathcal{M}_+(\mathtt{I}_{ij}(t)+\mathtt{X}_{ij}(t)+\mathtt{Y}_{ji}(t)) & $k_i>0>k_j$, $k_i\neq-k_j$\\
   \mathcal{M}_+( \mathtt{I}_{ij}(t)+\mathtt{Y}_{ij}(t)+\mathtt{X}_{ji}(t)) & $k_j>0>k_i$, $k_i\neq-k_j$\\
    \mathcal{M}^2_+(\mathtt{I}_{ij}(t)+\mathtt{Y}_{ij}(t)+\mathtt{Y}_{ji}(t)) & $k_i\neq k_j<0$\\
  \mathtt{I}_{ii}(t)+ \mathtt{D}_i(t) & $k_i=k_j>0$ \\
   \mathcal{M}_+(\mathtt{I}_{ii}(t)+\mathtt{D}_i(t)) & $k_i=-k_j>0$ \\
   \mathcal{M}_+(\mathtt{I}_{ii}(t)+\mathtt{E}_i(t)) & $k_i=-k_j<0$ \\
  \mathcal{M}^2_+(\mathtt{I}_{ii}(t)+\mathtt{E}_i(t)) &  $k_i=k_j<0$
  \end{dcases*}.
\end{equation}
Next we turn to $\mathtt{M}_-(k_i,k_j;t)$, which is much simpler.
Observe
\begin{align}
    \mathtt{L}^*_-(k;m_-)&=\ii\mathcal{N}^*_{\mathrm{L},-}\frac{\sqrt{N+2}}{2}(\vartheta(k)+\mathcal{M}_-\vartheta(-k))e^{-\ii m\delta_-(m_-)}(\delta_{k,m}-\delta_{k,-m})\n
    \mathtt{R}^*_-(k;m_-)&=\ii\mathcal{N}_{\mathrm{R},-}\frac{\sqrt{N+2}}{2}(\vartheta(k)+\mathcal{M}_-\vartheta(-k))e^{-\ii m\delta_-(m_-)}(\delta_{k,m}-\delta_{k,-m}).
\end{align}
Hence
\begin{equation}
    \mathtt{M}_-(k_i,k_j;t)=\frac{\gamma^2}{2\Tilde{\Gamma}}\begin{dcases*}
  1  & $k_i= k_j>0$ \\
  -\mathcal{M}_- &  $k_i=-k_j$\\
  \mathcal{M}_-^2 &  $k_i=k_j<0$ \\
  0 & otherwise
  \end{dcases*}.
\end{equation}
Combining everything, we finally obtain the full $\mathtt{M}(k_i,k_j;t)$, which turns out to be of the following form
\begin{equation}
    \mathtt{M}(k_i,k_j;t)=\delta_{k_i,k_j}e^{-\ii E_{k_i} t}+\frac{2\pi}{N+2}\mathtt{A}_{ij}(t).
\end{equation}
This implies that the Loschmidt echo has the form of the Fredholm determinant as expected
\begin{align}
    \mathcal{L}(t)&=\det_{i,j\in\mathrm{FS}}\left(1+\frac{2\pi}{N+2}\mathtt{B}(t)\right)_{ij}\n
    &\underset{N\to\infty}{\longrightarrow}\det(1+\mathtt{B}(t)),
\end{align}
where $\mathtt{B}_{ij}(t)=\left(\mathrm{diag}(e^{\ii E_{k_i}t})\mathtt{A}(t)\right)_{ij}$, and the determinant in the last line is performed over the $L^2$ space.
\subsection{Phase II}
When $\gamma_1^2>\sqrt{1+\gamma^2_2}$, the system acquires a new type of excitations, which are bound states that are localised across the dot. Labelling them with $m=\pm$, the corresponding overlap matrices read
\begin{align}
    \mathtt{L}^*(k,+)&=\frac{2(\mathcal{N}_{\mathrm{L},\mathrm{b}}^{\rm re})^*}{\sqrt{N+2}}\sum_{x>0}\left(\vartheta(k)+\frac{\gamma^2}{|\gamma|^2}\vartheta(-k)\right)\sin kx \,e^{-x/\xi} \n
     \mathtt{L}^*(k,-)&=\frac{2(\mathcal{N}_{\mathrm{L},\mathrm{b}}^{\rm re})^*}{\sqrt{N+2}}\sum_{x>0}\left(\vartheta(k)+\frac{\gamma^2}{|\gamma|^2}\vartheta(-k)\right)(-1)^x\sin kx \,e^{-x/\xi} \n
    \mathtt{R}(k,+)&=\frac{2\mathcal{N}_{\mathrm{R},\mathrm{b}}^{\rm re}}{\sqrt{N+2}}\sum_{x>0}\left(\vartheta(k)+\frac{\gamma^2}{|\gamma|^2}\vartheta(-k)\right)\sin kx\, e^{-x/\xi}\n
    \mathtt{R}(k,-)&=\frac{2\mathcal{N}_{\mathrm{R},\mathrm{b}}^{\rm re}}{\sqrt{N+2}}\sum_{x>0}\left(\vartheta(k)+\frac{\gamma^2}{|\gamma|^2}\vartheta(-k)\right)(-1)^x\sin kx\, e^{-x/\xi}.
\end{align}
Since
\begin{equation}
    \sum_{x>0}\sin kx\,e^{-x/\xi}=\frac{\sin k+ (e^{-1/\xi})^{N/2+1}\sin \frac{kN}{2}}{2(\cos 1/\xi-\sin k)}=\frac{\sqrt{2\tgamma-1}\sin k}{2(\tgamma-\sqrt{2\tgamma-1}\cos k)} + \mathcal{O}(e^{-N/(2\xi)}),
\end{equation}
and
\begin{equation}
    \sum_{x>0}(-1)^x\sin kx\,e^{-x/\xi}=-\frac{\sin k+ (-e^{-1/\xi})^{N/2+1}\sin \frac{kN}{2}}{2(\cos 1/\xi+\sin k)}=-\frac{\sqrt{2\tgamma-1}\sin k}{2(\tgamma+\sqrt{2\tgamma-1}\cos k)} + \mathcal{O}(e^{-N/(2\xi)}),
\end{equation}
we obtain
\begin{equation}
    \mathtt{L}^*(k,\pm)=\frac{(\mathcal{N}_{\mathrm{L},\mathrm{b}}^{\rm im})^*\sqrt{2\tgamma-1}}{\sqrt{N+2}}\mathbb{A}^\mathrm{re}_\pm(k),\quad \mathtt{R}(k,\pm)=\frac{\mathcal{N}_{\mathrm{R},\mathrm{b}}^{\rm im}\sqrt{2\tgamma-1}}{\sqrt{N+2}}\mathbb{A}^\mathrm{re}_\pm(k)
\end{equation}
with
\begin{equation}
    \mathbb{A}^\mathrm{re}_\pm(k)=\pm\left(\vartheta(k)+\frac{\gamma^2}{|\gamma|^2}\vartheta(-k)\right)\frac{\sin k}{\tgamma\mp\sqrt{2\tgamma-1}\cos k}.
\end{equation}
the additional contribution to the matrix $\mathtt{M}(k_i,k_j;t)$, which we denote $\mathtt{M}^\mathrm{re}_\mathrm{b}(k_i,k_j;t)$ reads
\begin{align}
    \mathtt{M}^\mathrm{re}_\mathrm{b}(k_i,k_j;t)&=\frac{(\gamma^*)^2(\Tilde{\Gamma}-1)}{N+2}\left(e^{-\ii E_\mathrm{b}t}\mathbb{A}^\mathrm{re}_+(k_i)\mathbb{A}^\mathrm{re}_+(k_j)+e^{\ii E_\mathrm{b}t}\mathbb{A}^\mathrm{re}_-(k_i)\mathbb{A}^\mathrm{re}_-(k_j)\right) + \mathcal{O}(e^{-N/(2\xi)})\n
    &=\frac{2\pi}{N+2} (\mathtt{A}^\mathrm{im}_\mathrm{b})_{ij}(t) + \mathcal{O}(e^{-N/(2\xi)}),
\end{align}
where
\begin{equation}
     (\mathtt{A}^\mathrm{re}_\mathrm{b})_{ij}(t)=\frac{(\gamma^*)^2(\Tilde{\Gamma}-1)}{2\pi}\left(e^{-\ii E_\mathrm{b}t}\mathbb{A}^\mathrm{im}_+(k_i)\mathbb{A}^\mathrm{im}_+(k_j)+e^{\ii E_\mathrm{b}t}\mathbb{A}^\mathrm{im}_-(k_i)\mathbb{A}^\mathrm{im}_-(k_j)\right).
\end{equation}
The Loschmidt echo in the Phase II can be then obtained by
\begin{align}
    \mathcal{L}(t)&=\det_{i,j\in\mathrm{FS}}\left(1+\frac{2\pi}{N+2}\left(\mathtt{B}(t)+\mathtt{B}^\mathrm{re}_\mathrm{b}(t)\right)\right)_{ij}\n
    &\underset{N\to\infty}{\longrightarrow}\det(1+\mathtt{B}(t)+\mathtt{B}^\mathrm{re}_\mathrm{b}(t)),
\end{align}
where
\begin{equation}\label{bre}
    (\mathtt{B}^\mathrm{re}_\mathrm{b})_{ij}(t)=\left(\mathrm{diag}(e^{\ii E_{k_i}t})\mathtt{A}^\mathrm{re}_\mathrm{b}(t)\right)_{ij}
\end{equation}
\subsection{Phase III}
In this regime, we have additional bound states with pure imaginary energy on top of the other particle-hole type modes. We can take them into account in the exactly the same way as in the phase II. What is different from the phase II is that the biorthogonal norm, which we also compute in this section, becomes time-dependent in this phase. Again labeling these two modes with $m=\pm$, the overlap matrices are given by
\begin{align}
    \mathtt{L}^*(k,\pm)&=\frac{2(\mathcal{N}_{\mathrm{L},\mathrm{b}}^{\rm im})^*}{\sqrt{N+2}}\left(\vartheta(k)+\frac{\gamma^2}{|\gamma|^2}\vartheta(-k)\right)\sum_{x>0}(\pm\ii)^x\sin kx \,e^{-x/\xi} \n
    \mathtt{R}(k,\pm)&=\frac{2\mathcal{N}_{\mathrm{R},\mathrm{b}}^{\rm im}}{\sqrt{N+2}}\left(\vartheta(k)+\frac{\gamma^2}{|\gamma|^2}\vartheta(-k)\right)\sum_{x>0}(\pm\ii)^x\sin kx \,e^{-x/\xi}.
\end{align}
Since
\begin{equation}
    \sum_{x>0}\ii^x\sin kx\,e^{-x/\xi}=\frac{\sin k+ \ii(e^{-1/\xi})^{N/2+1}\sin \frac{kN}{2}}{2(\sin 1/\xi-\sin k)}=\frac{\sqrt{1-2\tgamma}\sin k}{2(\ii\tgamma-\sqrt{1-2\tgamma}\cos k)}+ \mathcal{O}(e^{-N/(2\xi)})
\end{equation}
and
\begin{equation}
    \sum_{x>0}(-\ii)^x\sin kx\,e^{-x/\xi}=-\frac{\sqrt{1-2\tgamma}\sin k}{2(\ii\tgamma+\sqrt{1-2\tgamma}\cos k)}+ \mathcal{O}(e^{-N/(2\xi)}),
\end{equation}
we obtain
\begin{equation}
    \mathtt{L}^*(k,\pm)=\frac{(\mathcal{N}_{\mathrm{L},\mathrm{b}}^{\rm im})^*\sqrt{1-2\tgamma}}{\sqrt{N+2}}\mathbb{A}_\pm(k),\quad \mathtt{R}(k,\pm)=\frac{\mathcal{N}_{\mathrm{R},\mathrm{b}}^{\rm im}\sqrt{1-2\tgamma}}{\sqrt{N+2}}\mathbb{A}_\pm(k)
\end{equation}
with
\begin{equation}
    \mathbb{A}^\mathrm{im}_\pm(k)=\pm\left(\vartheta(k)+\frac{\gamma^2}{|\gamma|^2}\vartheta(-k)\right)\frac{\sin k}{2(\ii\tgamma\mp\sqrt{1-2\tgamma}\cos k)}.
\end{equation}
Notice that, in terms of $\Tilde{\Gamma}$, $\sin 1/\xi$ can also be expressed as $\sin 1/\xi=\ii(1-\Tilde{\Gamma})/\sqrt{1-2\Tilde{\Gamma}}$. The additional contribution to the matrix $\mathtt{M}(k_i,k_j;t)$, which we denote $\mathtt{M}^\mathrm{im}_\mathrm{b}(k_i,k_j;t)$ then reads
\begin{align}
    \mathtt{M}^\mathrm{im}_\mathrm{b}(k_i,k_j;t)&=\frac{(\gamma^*)^2(1-\Tilde{\Gamma})}{N+2}\left(e^{-E_\mathrm{b}t}\mathbb{A}^\mathrm{im}_+(k_i)\mathbb{A}^\mathrm{im}_+(k_j)+e^{E_\mathrm{b}t}\mathbb{A}^\mathrm{im}_-(k_i)\mathbb{A}^\mathrm{im}_-(k_j)\right) + \mathcal{O}(e^{-N/(2\xi)})\n
    &=\frac{2\pi}{N+2} (\mathtt{A}^\mathrm{im}_\mathrm{b})_{ij}(t) + \mathcal{O}(e^{-N/(2\xi)}),
\end{align}
where
\begin{equation}
     (\mathtt{A}^\mathrm{im}_\mathrm{b})_{ij}(t)=\frac{(\gamma^*)^2(1-\Tilde{\Gamma})}{2\pi}\left(e^{-E_\mathrm{b}t}\mathbb{A}^\mathrm{im}_+(k_i)\mathbb{A}^\mathrm{im}_+(k_j)+e^{E_\mathrm{b}t}\mathbb{A}^\mathrm{im}_-(k_i)\mathbb{A}^\mathrm{im}_-(k_j)\right).
\end{equation}
The Loschmidt echo in the Phase III can be computed as
\begin{align}
    \mathcal{L}(t)&=\det_{i,j\in\mathrm{FS}}\left(1+\frac{2\pi}{N+2}\left(\mathtt{B}(t)+\mathtt{B}^\mathrm{im}_\mathrm{b}(t)\right)\right)_{ij}\n
    &\underset{N\to\infty}{\longrightarrow}\det(1+\mathtt{B}(t)+\mathtt{B}^\mathrm{im}_\mathrm{b}(t)),
\end{align}
where
\begin{equation}\label{bim}
    (\mathtt{B}^\mathrm{im}_\mathrm{b})_{ij}(t)=\left(\mathrm{diag}(e^{\ii E_{k_i}t})\mathtt{A}^\mathrm{im}_\mathrm{b}(t)\right)_{ij}.
\end{equation}
Next we compute the biorthogonal norm of $|\psi(t)\rangle=e^{-\ii Ht}|\mathrm{GS}\rangle$. Thanks to the fact that $|\mathcal{N}_\mathrm{L}|^2=|\mathcal{N}_\mathrm{R}|^2$ \cite{biorth},we can express the norm $\langle\Tilde{\psi}(t)|\psi(t)\rangle$ as
\begin{equation}
    \langle\Tilde{\psi}(t)|\psi(t)\rangle=\det_{i,j\in\mathrm{FS}}\mathtt{N}(k_i;k_j),
\end{equation}
where
\begin{align}\label{binorm}
\mathtt{N}(k_i;k_j)&=\sum_me^{-\ii(E_m-E^*_m)t}\mathtt{L}^*(k_i,m)\mathtt{R}(k_j,m)\n
&=\delta_{k_i,k_j}+\sum_{\epsilon=\pm}\left(e^{-2\ii E^\mathrm{im}_{\mathrm{b},\epsilon}t}-1\right)\mathtt{L}^*(k_i,\epsilon)\mathtt{R}(k_j,\epsilon) \n
&=\delta_{k_i,k_j}+\frac{(\gamma^*)^2(1-\Tilde{\Gamma})}{N+2}\sum_{\epsilon=\pm}\left(e^{-2\epsilon E_{\mathrm{b}}t}-1\right)\mathbb{A}^\mathrm{im}_\epsilon(k_i)\mathbb{A}^\mathrm{im}_\epsilon(k_j)\n
&=:\delta_{k_i,k_j}+\frac{2\pi}{N+2}\Tilde{\mathtt{N}}(t)
\end{align}
which is again of the form of the Fredholm determinant. Therefore the norm is given by
\begin{equation}
    \langle\Tilde{\psi}(t)|\psi(t)\rangle=\det_{i,j\in\mathrm{FS}}\left(1+\frac{2\pi}{N+2}\Tilde{\mathtt{N}}(t)\right).
\end{equation}
\end{widetext}

\section{Details about the numerical simulations}
\subsection{Single particle energies and states}

\begin{figure}[h]
\includegraphics[width=0.5\textwidth]{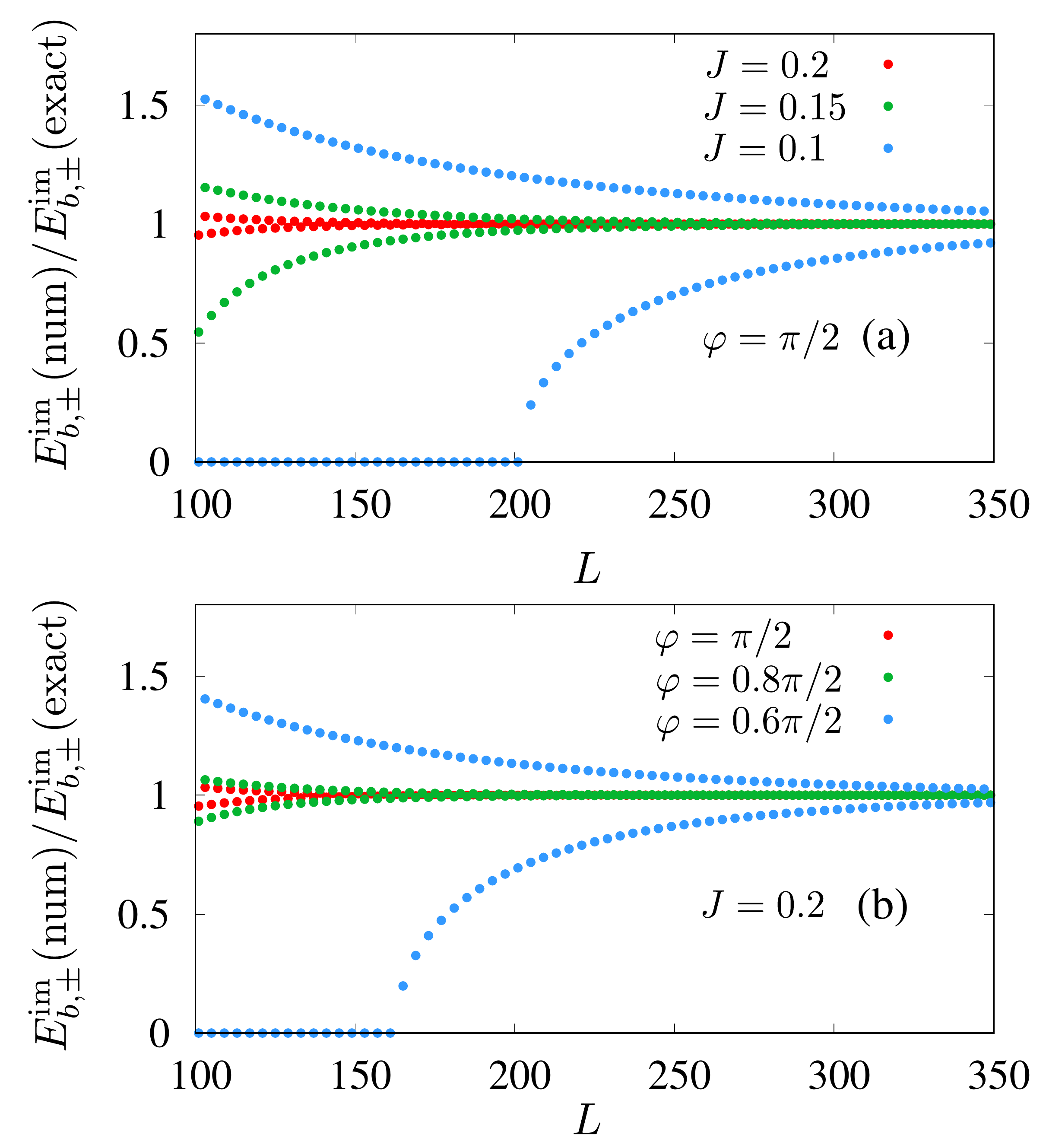}
\caption{Bound state energy $E^{\rm im}_{b,\pm}$ in the phase III as a function of system size $L$ for different values of a coupling magnitude $J$ [panel(a)] and of a coupling argument $\varphi$ [panel (b)]. For particular system sizes $L=5 \ \rm mod \ 4$ and $L< 2 \xi$ the single particle energies are real, so $E^{\rm im}_{b,\pm} = 0$.
}
\label{fig:En_vs_L}
\end{figure}

 A Hamiltonian Eq.\eqref{eq:ptrlm} of two leads coupled to the impurity can be diagonalized using a "Bogoliubov" transformation 
    \begin{eqnarray}
    H= \sum_{i,j} c^\dagger_i J_{ij}c_j = \sum_k \mathcal{E} _k  d^{\dagger}_{k R} d_{k L} \label{eq:H},\\
    d^\dagger_{k R}=\sum_j c^\dagger_j R_{jk},\  d_{k L}=\sum_j L_{kj} c_j  \\
    \sum_{ij} L_{ki}J_{ij}R_{jp}=\delta_{kp}\mathcal{E}_p,
\end{eqnarray}
where $\mathcal{E}_p$ are single excitation energies, $L$ and $R=L^{-1}$ are corresponding left and right eigenvectors of $J$. In the standard Hermitian case $J=J^\dagger$ a Bogoliubov transformation simplifies to a unitary one 
    \begin{eqnarray}
     H_0= \sum_{i,j} c^\dagger_i J_{ij}c_j = \sum_k \mathcal{E}^0_k  f^{\dagger}_{k} f_{k } \label{eq:H_0},\\
    f^\dagger_{k }=\sum_j c^\dagger_j U^\dagger_{jk},\  f_{k }=\sum_j U_{kj} c_j  \\
    \sum_{ij} U_{ki}J_{ij}U^\dagger_{jp}=\delta_{kp}\mathcal{E}^0_k.
    \end{eqnarray}
Single excitation energies $\mathcal{E}^0_k$ are real, whereas two of $\mathcal{E}_k$ are purely imaginary ($\mathcal{E}_\pm=E^{\rm im}_{\rm b,\pm}$) and correspond to the bound states in the phase III, see Fig. \ref{fig:En_Loc}. It turns out that for finite systems sizes $L=5 \ (\rm mod \ 4)$ the imaginary energies vanish $E^{\rm im}_{\rm b,\pm}({\rm num}) = 0$, as in Fig. \ref{fig:En_vs_L}. They arise for a system size $L > 2 \xi$ and  $E^{\rm im}_{\rm b,\pm}({\rm num})$ converge to the value given by Eq.\eqref{eq:imEb} in the thermodynamic limit $L \to \infty$.

\subsection{Two-point correlation functions}

\subsection{Loschmidt echo}
\subsubsection{Unnormalized echo $\mathcal{L}$(t)}
Recall that we define a normalized returning amplitude, or normalized Loschmidt echo, as 
\begin{equation}
  \bar{ \mathcal{L}}(t)=\frac{\mathcal{L}(t)}{\sqrt{|\langle\tilde{\psi}(t)|\psi(t)\rangle|}},
    \label{eq:echo0}
\end{equation} 
where   $\mathcal{L}=\bra{\psi_0} e^{-i H t} \ket{\psi_0}$. The initial state $\ket{\psi_0}$ is a ground state of a system $H_0(\gamma=0)$ of the impurity and two {\it disjoint} wires, a Hamiltonian $H$ describes wires coupled ($\gamma\neq 0$) to the dot. To calculate the time evolution, we notice that 
\begin{eqnarray}
    e^{-iHt}\ket{\psi_0}=e^{-iHt}f^\dagger_{k_1}...f^\dagger_{k_{N/2}}\ket{0}= \nonumber \\
    e^{-iHt}f^\dagger_{k_1}e^{i Ht}...e^{-iHt}f^\dagger_{k_{N/2}}e^{iHt}\ket{0}=\nonumber \\
    f^\dagger_{k_1}(t)...f^\dagger_{k_{N/2}}(t)\ket{0}=\ket{\psi(t)},
\end{eqnarray}
where each creation operator evolves according to 
\begin{eqnarray}
    f^\dagger_k(t)= e ^{-iHt}f^\dagger_k e^{iHt} = 
    \sum_{p} e^{-i Ht} d^\dagger_p (L U^\dagger)_{pk}  e^{i H t}= \nonumber \\
    \sum_{p} d^\dagger_p e^{-it \mathcal{E}_p}(L U^\dagger)_{pk}= \sum_{q} f^\dagger_q (U R e^{-it \mathcal{E}} L U^\dagger)_{qk}=\nonumber \\ \sum_q f^\dagger_q A(t)_{qk}.
\end{eqnarray}
Then the numerator of the Loschmidt echo is given by 
\begin{eqnarray}
    \mathcal{L}(t)=\la \psi_0| \psi(t)\ra = \nonumber \\
    \bra{0} f_{k_{N/2}}...f_{k_1} f^\dagger_{k_1}(t)... f^\dagger_{k_{N/2}}(t) \ket{0}= \nonumber \\ 
    \det \bra{0} f_k f^\dagger_{p}(t) \ket{0} =
    \det A(t)_{kp},
    \label{eq:det}
\end{eqnarray}
 indices run a Fermi sea $\mathcal{E}^0_{k,p}<0$, or, simply, we just eliminate rows in $U$, which correspond to positive singe excitation energies.

\subsubsection{Norm of a time evolving state}
\begin{figure}[h]
\includegraphics[width=0.5\textwidth]{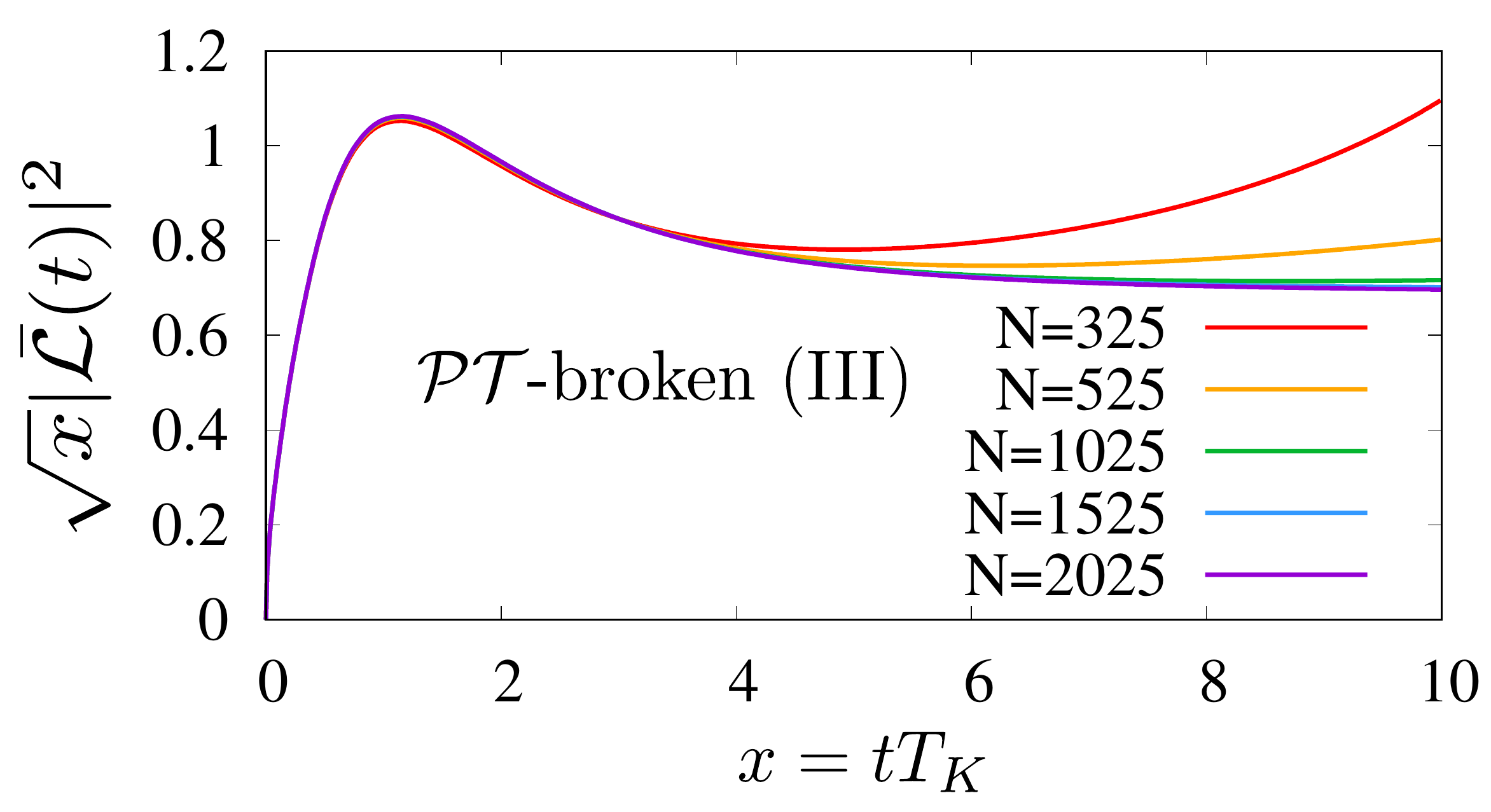}
\caption{Rescaled Loschmidt echo as a function of rescaled time for imaginary coupling $\gamma= 0.2 e^{i \pi/2}$. Scaling factor $T_K = |E^{\rm im}_b|$ Eq.\eqref{eq:imEb}. Color labels different values of a system size. }
\label{fig:Echo_N}
\end{figure}

According to the recipe of constructing a biorthogonal basis, in the conjugated state $\la\widetilde{\psi}(t)|$ we replace $R^\dagger$ by $L=R^{-1}$ and correspondingly $L^\dagger$ by $R=L^{-1}$. Then the $\mathcal{PT}$-norm of the time evolving state $\ket{\psi(t)}$ is
\begin{equation}
    \la \widetilde{\psi}(t)|\psi(t)\ra= \det (U R e^{-it (\mathcal{E}-\mathcal{E}^*)}LU^\dagger)_{kp},
\end{equation}
where indices run a Fermi sea $\mathcal{E}^0_{k,p}<0$, as in Eq.\eqref{eq:det}.  Numerical values for normalized Loschmidt echo converges for the large system sizes, see Fig. \ref{fig:Echo_N} 

\subsection{Analytical continuation of Loschmidt echo}
\label{ssec:coefficients}
We expand Eq.\eqref{eq:echo0} in Taylor series of the following form:
\begin{eqnarray}
\bar{\mathcal{L}}(t)= \sum_{n=0}a_n \gamma^n
\end{eqnarray}

In order to obtain coefficients $a_n$ from numerical data, we implement central finite difference operators (discrete derivative) of higher orders\cite{Fornberg} in the vicinity of $\gamma = 0$, which help to increase precision of calculations. For instance 
\begin{eqnarray}
hf'(x)_{x=0} =  \frac{1}{280} f(x - 4h)  +  \frac{-4}{105}f(x - 3h)  + 
   \frac{1}{5} f(x - 2h) \nonumber \\+    \frac{-4}{5}f(x - h) + \frac{4}{5} f(x + h) + \frac{-1}{5} f(x + 2h) \nonumber \\ + \frac{4}{105}f(x + 3h)  + \frac{-1}{280} f(x + 4h)+ \mathcal{O}(h^8), \qquad \quad
\end{eqnarray}
 where $h$ is a distance between neighboring points, which we chose $h=\Delta \gamma = 0.01$ and $\gamma \in (-0.2,0.2)$ . We want to pay attention that decreasing $h$ may lead to increasing errors, so one should choose an optimal $h$.

\end{document}